
\input phyzzx
\catcode`@=11
%
%

%
\font\fourteentt=cmtt10 scaled\magstep2  
\def\seventeenpoint{\relax
    \textfont0=\seventeenrm	    \scriptfont0=\twelverm
    \scriptscriptfont0=\tenrm
     \def\rm{\fam0 \seventeenrm \f@ntkey=0 }\relax
    \textfont1=\seventeeni	    \scriptfont1=\twelvei
    \scriptscriptfont1=\teni
     \def\oldstyle{\fam1 \seventeeni\f@ntkey=1 }\relax
    \textfont2=\seventeensy	    \scriptfont2=\twelvesy
    \scriptscriptfont2=\tensy
    \textfont3=\seventeenex     \scriptfont3=\seventeenex
    \scriptscriptfont3=\seventeenex
    \def\it{\fam\itfam \seventeenit\f@ntkey=4 }
         \textfont\itfam=\seventeenit
    \def\sl{\fam\slfam \seventeensl\f@ntkey=5 }
         \textfont\slfam=\seventeensl
    \scriptfont\slfam=\twelvesl
    \def\bf{\fam\bffam \seventeenbf\f@ntkey=6 }
         \textfont\bffam=\seventeenbf
    \scriptfont\bffam=\twelvebf	 \scriptscriptfont\bffam=\tenbf
    \def\tt{\fam\ttfam \fourteentt \f@ntkey=7 }
         \textfont\ttfam=\fourteentt
    \h@big=11.9\p@{} \h@Big=16.1\p@{} \h@bigg=20.3\p@{} \h@Bigg=24.5\p@{}
    \setbox\strutbox=\hbox{\vrule height 12pt depth 5pt width\z@}
    \samef@nt}

\font\twentyonebf=cmbx10 scaled \magstep4

%
%
%
%
\mathchardef\bfalpha  ="0B0B
\mathchardef\bfbeta   ="0B0C
\mathchardef\bfgamma  ="0B0D
\mathchardef\bfdelta  ="0B0E
\mathchardef\bfepsilon="0B0F
\mathchardef\bfzeta   ="0B10
\mathchardef\bfeta    ="0B11
\mathchardef\bftheta  ="0B12
\mathchardef\bfiota   ="0B13
\mathchardef\bfkappa  ="0B14
\mathchardef\bflambda ="0B15
\mathchardef\bfmu     ="0B16
\mathchardef\bfnu     ="0B17
\mathchardef\bfxi     ="0B18
\mathchardef\bfpi     ="0B19
\mathchardef\bfrho    ="0B1A
\mathchardef\bfsigma  ="0B1B
\mathchardef\bftau    ="0B1C
\mathchardef\bfupsilon="0B1D
\mathchardef\bfphi    ="0B1E
\mathchardef\bfchi    ="0B1F
\mathchardef\bfpsi    ="0B20
\mathchardef\bfomega  ="0B21
%
%
\newtoks\heth
\newtoks\Heth
\Pubnum={KUNS \the\pubnum}
\Heth={HE(TH)\the\heth}
\date={\monthname,\ \number\year}
\pubnum={000}
\heth={00/00}
\def\titlepage{\FRONTPAGE\ifPhysRev\PH@SR@V\fi
   \ifp@bblock\p@bblock\fi}
\def\p@bblock{\begingroup \tabskip=\hsize minus \hsize
   \baselineskip=1.5\ht\strutbox \topspace-2\baselineskip
   \halign to\hsize{\strut ##\hfil\tabskip=0pt\crcr
   \the\Pubnum\cr \the\Heth\cr \the\date\cr }\endgroup}
\def\titlestyleb#1{\par\begingroup \interlinepenalty=9999
     \leftskip=0.00\hsize plus 1.23\hsize minus 0.02\hsize
     \rightskip=\leftskip \parfillskip=0pt
     \hyphenpenalty=9000 \exhyphenpenalty=9000
     \tolerance=9999 \pretolerance=9000
     \spaceskip=0.333em \xspaceskip=0.5em
     \iftwelv@\fourteenpoint\else\twelvepoint\fi
   \noindent {\bf #1}\par\endgroup }
\def\title#1{\vskip\frontpageskip \titlestyleb{#1} \vskip\headskip }
%
%

%
%
\paperfootline={\hss\iffrontpage\else\ifp@genum%
		\tenrm --\thinspace\folio\thinspace --\hss\fi\fi}
%
%

%

%
%

%
\def\contr#1#2#3{\vbox{\ialign{##\crcr
	  \hskip #2pt\vrule depth 4pt
          \hrulefill\vrule depth 4pt\hskip #3pt
	  \crcr\noalign{\kern-1pt\vskip0.125cm\nointerlineskip}
          $\hfil\displaystyle{#1}\hfil$\crcr}}}
\def\leftrightarrowfill{$\m@th\mathord-\mkern-6mu%
  \cleaders\hbox{$\mkern-2mu\mathord-\mkern-2mu$}\hfill
  \mkern-6mu\mathord\leftrightarrow$}
\def\overleftrightarrow#1{\vbox{\ialign{##\crcr
      \leftrightarrowfill\crcr\noalign{\kern-\p@\nointerlineskip}
      $\hfil\displaystyle{#1}\hfil$\crcr}}}
%

%
%

%
\def\rbox#1{\vbox{\hrule height.8pt%
		\hbox{\vrule width.8pt\kern5pt
		\vbox{\kern5pt\hbox{#1}\kern5pt}\kern5pt
		\vrule width.8pt}
		\hrule height.8pt}}
%
%
\def\sqr#1#2{{\vcenter{\hrule height.#2pt
      \hbox{\vrule width.#2pt height#1pt \kern#1pt
          \vrule width.#2pt}
      \hrule height.#2pt}}}
\def\overbar#1{\vbox{\ialign{##\crcr
	  \hskip 1.5pt\hrulefill\hskip 1.1pt
	  \crcr\noalign{\kern-1pt\vskip0.125cm\nointerlineskip}
          $\hfil\displaystyle{#1}\hfil$\crcr}}}
%
%
%

%
%

%
%
\mathchardef\Lag="724C
%
%

%

%
\def\addeqno{\ifnum\equanumber<0 \global\advance\equanumber by -1
    \else \global\advance\equanumber by 1\fi}
\def\undereq#1{\mathop{\vtop{\ialign{##\crcr
      $\hfil\displaystyle{#1}\hfil$
      \crcr\noalign{\kern3\p@\nointerlineskip}
      \crcr\noalign{\kern3\p@}}}}\limits}
\def\overeq#1{\mathop{\vbox{\ialign{##\crcr\noalign{\kern3\p@}
      \crcr\noalign{\kern3\p@\nointerlineskip}
      $\hfil\displaystyle{#1}\hfil$\crcr}}}\limits}
%

%
%

%
%
\def\journal#1&#2(#3){\unskip, {\sl #1}{\bf #2}(19#3)}
\def\andjournal#1&#2(#3){{\sl #1}{\bf #2}(19#3)}
\def\andvol&#1(#2){{\bf #1}(19#2)}

\def\NP{Nucl. Phys. }
\def\PR{Phys. Rev. }
\def\PRL{Phys. Rev. Lett. }
\def\PL{Phys. Lett. }
\def\PTP{Prog. Theor. Phys. }

%
%
\def\acknowledge{\par\penalty-100\medskip \spacecheck\sectionminspace
   \line{\hfil ACKNOWLEDGEMENTS\hfil}\nobreak\vskip\headskip }
%
%

%
%

%
\catcode`@=12
%
%

%

%


\catcode`@=11 

\def\p@bblock{\begingroup \tabskip=\hsize minus \hsize
   \baselineskip=1.5\ht\strutbox \topspace-2\baselineskip
   \halign to\hsize{\strut ##\hfil\tabskip=0pt\crcr
   \the\Pubnum\cr \the\Heth\cr \the\date\cr
   \the\pubmemo\cr 
   }\endgroup}

\def\datenum{\number\month /\number\day}
\newtoks\pubmemo
\pubmemo={\twentyonebf Ver.  \datenum}

\catcode`@=12


\footline={\hfill\ -- \folio\ -- \hfill}
\def\prenum#1{\rightline{#1}}
\def\date#1{\rightline{#1}}
\def\IJMP{Int. J. Mod. Phys.}

%
\catcode`@=11
%
%
\font\fourteenmib=cmmib10 scaled\magstep2   \skewchar\fourteenmib='177
\font\twelvemib=cmmib10 scaled\magstep1     \skewchar\twelvemib='177
\font\elevenmib=cmmib10 scaled\magstephalf  \skewchar\elevenmib='177
\font\tenmib=cmmib10                        \skewchar\tenmib='177
%
\font\fourteenbsy=cmbsy10 scaled\magstep2   \skewchar\fourteenbsy='60
\font\twelvebsy=cmbsy10 scaled\magstep1     \skewchar\twelvebsy='60
\font\elevenbsy=cmbsy10 scaled\magstephalf  \skewchar\elevenbsy='60
\font\tenbsy=cmbsy10                        \skewchar\tenbsy='60
%
%
\newfam\mibfam
%
%
\def\fourteenf@nts{\relax
    \textfont0=\fourteenrm          \scriptfont0=\tenrm
      \scriptscriptfont0=\sevenrm
    \textfont1=\fourteeni           \scriptfont1=\teni
      \scriptscriptfont1=\seveni
    \textfont2=\fourteensy          \scriptfont2=\tensy
      \scriptscriptfont2=\sevensy
    \textfont3=\fourteenex          \scriptfont3=\twelveex
      \scriptscriptfont3=\tenex
    \textfont\itfam=\fourteenit     \scriptfont\itfam=\tenit
    \textfont\slfam=\fourteensl     \scriptfont\slfam=\tensl
    \textfont\bffam=\fourteenbf     \scriptfont\bffam=\tenbf
      \scriptscriptfont\bffam=\sevenbf
    \textfont\ttfam=\fourteentt
    \textfont\cpfam=\fourteencp
    \textfont\mibfam=\fourteenmib   \scriptfont\mibfam=\tenmib
    \scriptscriptfont\mibfam=\tenmib }
\def\twelvef@nts{\relax
    \textfont0=\twelverm          \scriptfont0=\ninerm
      \scriptscriptfont0=\sixrm
    \textfont1=\twelvei           \scriptfont1=\ninei
      \scriptscriptfont1=\sixi
    \textfont2=\twelvesy           \scriptfont2=\ninesy
      \scriptscriptfont2=\sixsy
    \textfont3=\twelveex          \scriptfont3=\tenex
      \scriptscriptfont3=\tenex
    \textfont\itfam=\twelveit     \scriptfont\itfam=\nineit
    \textfont\slfam=\twelvesl     \scriptfont\slfam=\ninesl
    \textfont\bffam=\twelvebf     \scriptfont\bffam=\ninebf
      \scriptscriptfont\bffam=\sixbf
    \textfont\ttfam=\twelvett
    \textfont\cpfam=\twelvecp
    \textfont\mibfam=\twelvemib   \scriptfont\mibfam=\tenmib
    \scriptscriptfont\mibfam=\tenmib }
\def\tenf@nts{\relax
    \textfont0=\tenrm          \scriptfont0=\sevenrm
      \scriptscriptfont0=\fiverm
    \textfont1=\teni           \scriptfont1=\seveni
      \scriptscriptfont1=\fivei
    \textfont2=\tensy          \scriptfont2=\sevensy
      \scriptscriptfont2=\fivesy
    \textfont3=\tenex          \scriptfont3=\tenex
      \scriptscriptfont3=\tenex
    \textfont\itfam=\tenit     \scriptfont\itfam=\seveni  
    \textfont\slfam=\tensl     \scriptfont\slfam=\sevenrm 
    \textfont\bffam=\tenbf     \scriptfont\bffam=\sevenbf
      \scriptscriptfont\bffam=\fivebf
    \textfont\ttfam=\tentt
    \textfont\cpfam=\tencp
    \textfont\mibfam=\tenmib   \scriptfont\mibfam=\tenmib
    \scriptscriptfont\mibfam=\tenmib }
\def\mib{\n@expand\f@m\mibfam}

%
%
\Twelvepoint
\catcode`@=12
%


\REF\Wes{for example, J.~Wess and J.~Bagger, {\sl Supersymmetry
	 and Supergravity} (Princeton University Press, 1983).}

\REF\Hir{D.~Bernard \journal Suppl.~Progr.~Theor.~Phys. &102 (90) 49;
	\nextline
	I.~Ya.~Aref'eva and I.~V.~Volovich \journal Mod.~Phys.~Lett. &A6
	(91) 893; \nextline
	M.~Hirayama \journal \PTP &88 (92) 111.
	}

\REF\Bou{for review,
	P.~Bouwknegt and K.~Schoutens \journal Phys.~Rep. &223 (93) 183.}

\REF\Uti{R.~Utiyama \journal \PR &101 (56) 1597}

\REF\Yan{C.~N.~Yang and R.~I.~Mills \journal \PR &96 (54) 191.}

\REF\Sch{K.~Schoutens, A.~Sevrin and P.~van Nieuwenhuizen
         \journal Commun.~Math.~Phys. &124 (89) 87;
         \andjournal Int.~J.~Mod.~Phys. &A6 (91) 2891;
         \andjournal \PL &B255 (91) 549.}

\REF\Wit{E.~Witten \journal \NP &B311 (88/89) 46;
         \andvol &B323 (89) 113.}

\REF\Jac{C.~Teitelboim \journal \PL &B126 (83) 41;
         in {\sl Quantum Theory of Gravity}, \nextline
         ed.~S.M.~Christensen (Adam Hilger, 1984); \nextline
         R.~Jackiw, in {\sl Quantum Theory of Gravity},
         ed.~S.M.~Christensen (Adam Hilger, 1984);
         \andjournal \NP &B252 (85) 343; \nextline
         M.~Henneaux \journal \PRL &54 (85) 959; \nextline
         See also M.~Abe and N.~Nakanishi \journal Int.~J.~Mod.~Phys.
         &A6 (91) 3955.}

\REF\Fuk{T.~Fukuyama and K.~Kamimura \journal \PL &B160 (85) 259;
         \nextline K.~Isler and C.A.~Trugenberger \journal \PRL
         &63 (89) 834; \nextline
         A.H.~Chamseddine and D.~Wyler \journal \PL &B228 (89) 75;
         \andjournal \NP &B340 (90) 595; \nextline
         H.~Terao, \journal \NP & B395 (93) 623; \nextline
         See also D.~Cangemi and R.~Jackiw \journal \PRL &69 (92) 233.}

\REF\Ike{N.~Ikeda and Izawa K.-I. \journal \PTP &89 (93) 223.}

\REF\Iza{N.~Ikeda and Izawa K.-I. \journal \PTP &89 (93) 1077.}

\REF\Dil{N.~Ikeda and Izawa K.-I. \journal \PTP &90 (93) 237.}

\REF\Sug{N.~Ikeda {\sl Int.~J.~Mod.~Phys.} {\bf A}, to be published.}

\REF\Zam{A.B.~Zamolodchikov \journal Theor.~Math. Phys. &65 (86) 1205.}

\REF\Dri{V.~G.~Drinfel'd, in {\sl Proceedings of International Congress
	of Mathematicians} (Berkeley, CA, USA, 1986) 793; \nextline
	M.~Jimbo \journal Lett.~Math.~Phys. &10 (85) 63.}

\REF\Roc{M.~Ro\v{c}ek \journal \PL &B255 (91) 554; \nextline
	T.~Tjin \journal \PL &B292 (92) 60; \nextline
	J.~de Boer and T.~Tjin, preprint THU-92-32.}

\REF\Bat{I.A.~Batalin and G.A.~Vilkovisky \journal \PL &B102 (81) 27;
         \andjournal \PR &D28 (83) 2567;
         \andjournal J.~Math.~Phys. &26 (85) 172.}

\REF\Kug{T.~Kugo and I.~Ojima \journal \PL &B73 (78) 459;
         \andjournal Suppl.~\PTP &66 (79) 1; \nextline
         M.~Kato and K.~Ogawa \journal \NP &B212 (83) 443; \nextline
         N.~Nakanishi and I.~Ojima, {\sl Covariant Operator
         Formalism of Gauge Theories and Quantum Gravity}
         (World Scientific, 1990).}

\REF\Kat{M.O.~Katanaev and I.V.~Volovich
         \journal JETP~Lett. &43 (86) 267;
         \andjournal
         \PL &B175 (86) 413; \andjournal Ann.~of~Phys. &197 (90) 1;
         \nextline
         M.O.~Katanaev \journal Theor.~Math.~Phys. &80 (89) 838;
         \andjournal Sov.~Phys.~Dokl. &34 (89) 982;
         \andjournal J.~Math.~Phys. &31 (90) 882; \andvol &32 (91) 2483;
         \nextline
         K.G.~Akdeniz, A.~Kizilers{\" u}, and E.~Rizao{\v g}lu
         \journal \PL &B215 (88) 81;
         \nextline
         K.G.~Akdeniz, {\" O}.F.~Dayi, and A.~Kizilers{\" u}
         \journal Mod.~Phys.~Lett. &A7 (92) 1757;
         \nextline
         W.~Kummer and D.J.~Schwarz \journal \PR &D45 (92) 3628;
         \andjournal \NP &B382 (92) 171; \nextline
         H.~Grosse, W.~Kummer, P.~Pre{\v s}najder, and D.J.~Schwarz
         \journal J.~Math.~Phys. &33 (92) 3892; \nextline
	 T.~Strobl, \journal \IJMP &A8 (93) 1383; \nextline
         P.~Schaller and T.~Strobl, preprint TUW-92-13, TUW-93-08; \nextline
	 H.~Haider and W.~Kummer, preprint TUW-92-15}

\REF\Yon{T.~Yoneya \journal \PL &B149 (84) 111; \nextline
         A.M.~Polyakov \journal Mod.~Phys.~Lett. &A2 (87) 893.}

\REF\Str{T.~Strobl \journal \IJMP &A8 (93) 1383}

\REF\Rus{J.G.~Russo and A.A.~Tseytlin \journal \NP &B382
	 (92) 259.}

\REF\Fay{P.~Fayet and S.~Ferrara \journal Phys.~Rep. &32C (77) 249}

\REF\How{P.~S.~Howe \journal J.~phys.~A:Math.~Gen. &12 (79) 393.
	}

\REF\Noj{S.~Nojiri and I.~Oda \journal Mod.~Phys.~Lett. &A8 (93) 53,
	\nextline
	Y.~Park and A.~Strominger \journal Phys.~Rev. &D47 (93) 1569,
	\nextline
	A.~Bilal \journal Phys.~Rev. &D48 (93) 1665.}

\REF\Ver{H.~Verlinde, in {\sl The Sixth Marcel Grossmann
         Meeting on General Relativity}, \nextline
         ed.~M.~Sato (World Scientific, Singapore, 1992); \nextline
         D.~Cangemi and R.~Jackiw \journal \PRL &69 (92) 233.}

\REF\Riv{V.~O.~Rivelles preprint IFUSP-P-1025 (1992).}


\footline={\hfil}



\prenum{RIMS-953}
\date{December 1993}
\vskip 1.5cm


\title{Two-Dimensional Gravity \break
       and Nonlinear Gauge Theory}

\author{Noriaki Ikeda}

\address{Research Institute for Mathematical Sciences \break
            Kyoto University, Kyoto 606-01, Japan}

\abstract{
We construct a gauge theory based on nonlinear Lie algebras,
which is an extension of the usual gauge theory based on Lie algebras.
It is a new approach to generalization of the gauge theory.
The two-dimensional gravity is derived from nonlinear
Poincar{\' e} algebra, which is the new Yang-Mills like formulation
of the gravitational theory.
As typical examples, we investigate $R^2$ gravity
with dynamical torsion and generic form of 'dilaton' gravity.
The supersymmetric extension of this theory is also discussed.
}

\endpage
\pageno=2
\footline{\hfill-\ \folio \ -\hfill}




\font\sc=cmr5 scaled\magstep1
\def\bold#1{#1\llap{$#1$\hskip.3pt}\llap{$#1$\hskip.4pt}\llap{$#1$\hskip.5pt}}
\def\gct{\delta_{\hbox {\sc G}}}
\def\iso{\delta}
\def\brssub{{\bold\delta}}
\def\brs{{{\bold\delta}_{\hbox {\sc B}}}}
\def\brso{{{\bold\delta}_{\hbox {\sc b}}}}
\def\susy{\delta_{\hbox {\sc S}}}
\def\bt{{\bar\theta}}
\def\cu{{\cal U}}
\def\cv{{\cal V}}
\def\wo{{\cal W}}
\def\wt{{\cal U}}
\def\calf{{\cal Z}}
\def\lnormal{{\>:\,}}
\def\rnormal{{\,:\>}}
\def\scdots{\!\!\!\!\cdots\!\!\!\!}

%
\chapter{Introduction}

The gauge theory is fundamental object
in quantum field theory.
In this paper, we consider a generalization
of the usual gauge theory based on Lie algebra and
its application.

There are many studies to extend it
by modifying symmetry structure,
such as
super\-symmetry\rlap,\refmark{\Wes}
quantum group\rlap,\refmark{\Hir} W algebra\rlap.\refmark{\Bou}
However in this paper,
we consider quite new approach to
extend the usual gauge theory.
%
We construct
a two-dimensional gauge theory based on
nonlinear extension of
Lie algebras as a generalization
of the usual nonabelian gauge theory with internal gauge symmetry.
We call it nonlinear gauge theory,
which is a new generalization of nonabelian gauge theory.

Nonlinear extension of Lie algebras which is called
nonlinear Lie algebra has
the following commutation relation:
$$
  [T_A, T_B] = W_{AB}(T),
$$
where $W_{AB}(T)$ denotes a polynomial in $T_C$.
The details of definition are presented in section 2.2.
Quadratic nonlinear Lie algebras in the context of quantum field theory
were first introduced by
K.~Schoutens, A.~Sevrin and P.~van Nieuwenhuizen\rlap.\refmark{\Sch}
They mainly analyzed the W-gravity as an
application of nonlinear Lie algebra to quantum field theory.
Our nonlinear gauge theory is
different realization of nonlinear Lie algebra
in quantum field theory, and presents a new category
of constrainted systems.

The two-dimensional nonlinear gauge theory is closely related to
two-dimensional gravity.
The investigation of diffeomorphism-invariant field theories
is expected to provide indispensable information for
the quest of quantum gravity.
At the classical level, the Einstein theory
--- field theory of metric tensor with general covariance ---
is quite successful in describing gravitational phenomena.

The general covariance in gravitation theory
is external gauge symmetry, which is spacetime symmetry.
It may be called gauge symmetry of the Utiyama type\rlap.\refmark{\Uti}
On the other hand, the internal symmetry such as in QED or QCD
is called the one of the Yang-Mills type\rlap.\refmark{\Yan}
The connection to gauge symmetry
of the Utiyama type with the Yang-Mills type
has been investigated on various authors.
Witten's three-dimensional Einstein
gravity (Chern-Simons-Witten gravity)
can be regarded as one example
for the above description\rlap.\refmark{\Wit}
In two dimension, Jackiw-Teitelboim's model\refmark{\Jac} is
a typical example that
a gravitational theory and a gauge theory
have classically the same action\rlap.\refmark{\Fuk}
One of the reasons
for investigating the connection between two types of symmetries
is that symmetry of the Yang-Mills type is easier
to treat in quantization problems than that of the Utiyama type.
In this paper, we introduce new examples which have such properties
and are related to nonlinear gauge theory.

The particular feature of two-dimensional nonlinear gauge theory is
that it includes two-dimensional gravitation theory.
When the nonlinear algebra is Lorentz-covariant extension of
the Poincar\'e algebra, the theory turns out to be
the Yang-Mills-like formulation of
$R^2$ gravity with dynamical torsion\rlap,
\refmark{\Ike}\refmark{\Iza} or generic form of `dilaton' gravity\rlap.
\refmark{\Dil}
The hidden symmetry of their gravitation theory is clarified.
It yields new examples of the connection between the
two types of gauge symmetries.
Moreover a gauge theory of nonlinear Lie superalgebra is related to
supergravity.
When nonlinear superalgebra is nonlinear and Lorentz-covariant
extension of super-Poincar\'e algebra, the theory turn out to
be dilaton supergravity\rlap.\refmark{\Sug}


The remaining part of the paper is organized as follows:
In chapter 2, our action for `nonlinear' gauge theory is constructed
in quite a general manner. We also consider nonlinear Lie algebras
as a background structure for the `nonlinear' gauge theory.
The BRS formulation of nonlinear gauge theory is formulated.
%
The chapter 3 provides Lorentz-covariant quadratic extension
of the Poincar{\' e} algebra and our nonlinear gauge theory
based on it.
It turn out that the resulting theory is the Yang-Mills-like
formulation of $R^2$ gravity with dynamical torsion.
In chapter 4,
the relevance of this `nonlinear' gauge theory to the `dilaton' gravity
is clarified.
In chapter 5, our action for `nonlinear' gauge theory is extended to
the supergauge theory.
The connection to the dilaton supergravity of 'nonlinear' supergauge theory
is analyzed.
%
Chapter 6 concludes the paper.


%
\chapter{`Nonlinear' Gauge Theory}

In this chapter, we explain a generic approach to the
construction of a `nonlinear' gauge theory
--- gauge theory based on nonlinear extension of
Lie algebras\rlap.\refmark{\Dil}

%
\section{`Nonlinear' Gauge Transformation}
%

We introduce a `coadjoint' scalar field $\Phi_A$ in addition to
a vector field $h_\mu^A$ for the purpose of
constructing a `nonlinear' gauge theory.
Here $A$ denotes an internal index.
We consider
the following extension of gauge transformation:
$$
 \eqalign{
  \delta (c) h_\mu^A
   &= \partial_\mu c^A + U_{BC}^A(\Phi) h_\mu^B c^C, \cr
  \delta (c) \Phi_A &= W_{BA}(\Phi) c^B, \cr
 }
 \eqn\GGAUGE
$$
where $\delta (c)$ is a gauge transformation with the gauge-transformation
parameter $c^A$, and
$U_{BC}^A(\Phi)$ and $W_{AB}(\Phi)$ are smooth functions of the field
$\Phi_A$ to be determined below.
In the case of usual nonabelian gauge theory,
$U^A_{BC} = f^A_{BC}$ and $W_{BA} = f^C_{BA} \Phi_C$,
where $f^C_{AB}$ denote structure constants in some Lie algebra.
Since the above gauge transformation should reduce to
the usual nonabelian one if
$U^A_{BC}$ is constant and  $W_{BA}$ is a linear function
of $\Phi_C$, we assume $U^A_{BC} = - U^A_{CB}$
and $W_{BA} = - W_{AB}$.

We investigate the commutator algebra of these gauge
transformations
with gauge-transformation parameters $c_1$ and $c_2$.
We require a closed algebra
on the scalar field $\Phi_A$
$$
  [\delta (c_1), \delta (c_2)]\Phi _A = \delta (c'_3)\Phi _A,
 \eqn\FCOMM
$$
First we consider the case that the gauge parameters
$c_1$ and $c_2$ are independent of $\Phi_A$.
In this case, the right-hand side of \FCOMM\ becomes
$$
 [\delta (c_1), \delta (c_2)]\Phi _A =
	\left( {\partial W_{CA} \over \partial \Phi_D} W_{BD}
	+ {\partial W_{AB} \over \partial \Phi_D}
	W_{CD} \right) c_1^B c_2^C.
\eqn\FCOMC
$$
In case of nonabelian gauge theory,
since $W_{BA} = f^C_{BA} \Phi_C$,
we can rewrite \FCOMC\ as
$$\eqalign{
 [\delta (c_1), \delta (c_2)]\Phi _A & =
	( f^D_{CA} f^E_{BD}
	+ f^D_{AB} f^E_{CD} ) \Phi_E c_1^B c_2^C, \cr
	& = f^D_{BC} f^E_{DA} \Phi_E c_1^B c_2^C, \cr
	& = \delta(c'_3) \Phi_A, \cr
}
\eqn\FCOMNABEL
$$
by using the Jacobi identity of Lie algebra:
$$
	f^D_{[AB} f^E_{C]D} = 0,
\eqn\LIEJAC
$$
where $[ABC]$ denotes antisymmetrization
in the indices $A$, $B$, and $C$,
and
$$
  c_3^A = f^A_{BC} c_1^Bc_2^C.
$$
In order to extend the above construction to \FCOMC,
we replace the structure constant
$f^C_{BA}$ by ${\partial W_{BA} \over \partial \Phi_C}$
and $f^C_{BA} \Phi_C$ by $W_{BA}$
in \FCOMNABEL.
Then \FCOMNABEL\ becomes \FCOMC,
and \LIEJAC\ is generalized to
$$
  { \partial W_{[AB} \over \partial \Phi_D }
   W_{C]D} = 0.
 \eqn\JACPHI
$$
Owing to this generalized Jacobi identity, \FCOMC\ becomes
$$\eqalign{
 [\delta (c_1), \delta (c_2)]\Phi _A & =
	\left( {\partial W_{CA} \over \partial \Phi_D} W_{BD}
	+ {\partial W_{AB} \over \partial \Phi_D}
	W_{CD} \right) c_1^B c_2^C \cr
	& = W_{DA} {\partial W_{BC} \over \partial \Phi_D } c_1^B c_2^C. \cr
}
\eqn\FCOMCOMP
$$
This assumption results in the composite parameter
$$
  c'^A_3 = {\partial W_{BC} \over \partial \Phi_A} c_1^B c_2^C,
 \eqn\COMPPP
$$
and the gauge algebra becomes the required form \FCOMM\ on $\Phi_A$.
In the next section, we show that the expression \JACPHI\
can be derived as the Jacobi identity for a nonlinear Lie algebra
with structure functions $W_{AB}$.

Since the right-hand side of \COMPPP\ involves $\Phi_A$,
we have to
consider the case that the gauge parameters
$c_4$ and $c_5$ are dependent on $\Phi_A$ in
$[\delta (c_4), \delta (c_5)]\Phi _A$.
In this case, we have
$$\eqalign{
 [\delta (c_4), \delta (c_5)]\Phi _A & =
        \left( {\partial W_{CA} \over \partial \Phi_D} W_{BD}
        + {\partial W_{AB} \over \partial \Phi_D}
        W_{CD} \right) c_4^B c_5^C \cr
	& \quad - \left[ W_{DA} W_{BC}
	\left( {\partial c_5^D \over \partial \Phi_B}
	c_4^C - {\partial c_4^D \over \partial \Phi_B}
	c_5^C \right) \right] \cr
        & = W_{DA} \left[ {\partial W_{BC} \over \partial \Phi_D }
	c_4^B c_5^C
	- W_{BC} \left( {\partial c_5^D \over \partial \Phi_B}
	c_4^C - {\partial c_4^D \over \partial \Phi_B}
	c_5^C \right) \right], \cr
}
\eqn\FCOMDEP
$$
by using \JACPHI, and if we set
$$
  c'^A_6 = {\partial W_{BC} \over \partial \Phi_A} c_4^B c_5^C
	- W_{BC} \left( {\partial c_5^A \over \partial \Phi_B}
        c_4^C - {\partial c_4^A \over \partial \Phi_B} c_5^C \right),
$$
the gauge algebra satisfies
$$
 [\delta (c_4), \delta (c_5)]\Phi _A = \delta (c^\prime_6) \Phi_A.
\eqn\GENFCOMM
$$

On the other hand,
as the commutator of two transformations
on the vector field $h_\mu^A$, we require
$$
  [\delta (c_1), \delta (c_2)]h_\mu^A = \delta (c_3)h_\mu^A + \cdots,
 \eqn\HCOMM
$$
where the ellipsis
indicates a term which is irrelevant here.
If we calculate the right-hand side of \HCOMM\ explicitly,
it is obtained that
$$\eqalign{
  [\delta (c_1), \delta (c_2)]h_\mu^A & =
	\partial_\mu ( U_{BC}^A c_1^B c_2^C )
	- {\partial U^A_{BC} \over \partial \Phi_D } \partial_\mu \Phi_D \cr
	& \quad + \left( U_{DC}^A U_{BE}^D + U_{DE}^A U_{CB}^D
	+ {\partial U_{BC}^A \over \partial \Phi_D} W_{ED}
	+ {\partial U_{EB}^A \over \partial \Phi_D} W_{CD} \right)
	h_\mu{}^B c_1^E c_2^C, \cr
}
\eqn\HCOMMCAL
$$
where we assume that $c_1$ and $c_2$ are independent
of $\Phi_A$.
{}From the first term of left-hand side of \HCOMMCAL,
the composite parameter $c_3$ should be
$$
  c_3^A = U^A_{BC} c_1^Bc_2^C.
 \eqn\COMPP
$$

We require $c_3 = c'_3$, that is, we henceforth put
$$
  U^A_{BC}( \Phi) =  {\partial W_{BC} \over \partial \Phi_A}.
 \eqn\UANDW
$$
in order for the commutator algebra to represent
a consistent composition law of Lagrangian symmetry.
Then \HCOMMCAL\ becomes
$$\eqalign{
  [\delta (c_1), \delta (c_2)]h_\mu^A & =
        \partial_\mu \left( {
	\partial W_{BC} \over \partial \Phi_A} c_1^B c_2^C \right)
        - {\partial^2 W_{BC} \over \partial \Phi_A \partial \Phi_D }
	\partial_\mu \Phi_D \cr
        & \hskip-2cm + \left( {\partial W_{DC} \over \partial \Phi_A }
	{ \partial W_{BE} \over \partial \Phi_D }
	+ {\partial W_{DE} \over \partial \Phi_A }
	{ \partial W_{CB} \over \partial \Phi_D }
        + {\partial^2 W_{BC} \over \partial \Phi_A \partial \Phi_D} W_{ED}
        + {\partial^2 W_{EB} \over \partial \Phi_A \partial \Phi_D} W_{CD}
	\right)
        h_\mu{}^B c_1^E c_2^C \cr
	& = \partial_\mu \left( {\partial W_{BC} \over  \partial
	\Phi_A} c_1^B c_2^C \right)
	+ { \partial W_{BD} \over \partial \Phi_A} h_\mu{}^B
	{ \partial W_{EC} \over \partial \Phi_D } c_1^E c_2^C
	- c_1^Cc_2^D { \partial^2 W_{CD} \over
        \partial \Phi_A \partial \Phi_B } D_\mu \Phi _B \cr
	& = \delta (c_3)h_\mu ^A
             - c_1^Cc_2^D { \partial^2 W_{CD} \over
                \partial \Phi_A \partial \Phi_B } D_\mu \Phi _B, \cr
}
\eqn\HCOMRUS
$$
by using \JACPHI.

In the case that $c_4$ and $c_5$ is dependent on $\Phi_A$,
the commutator of two transformations on $h_\mu{}^A$ is
calculated by using \JACPHI\ as follows:
$$\eqalign{
  &[\delta (c_4), \delta (c_5)]h_\mu ^A \,= \delta (c_6)h_\mu ^A
             - c_4^Cc_5^D { \partial^2 W_{CD} \over
                \partial \Phi_A \partial \Phi_B } D_\mu \Phi _B, \cr
}
$$
where
$$
  c^A_6 = {\partial W_{BC} \over \partial \Phi_A} c_4^B c_5^C
        + W_{BC} \left( {\partial c_5^D \over \partial \Phi_B}
        c_4^C - {\partial c_4^D \over \partial \Phi_B} c_5^C \right),
$$
which is the same as \GENFCOMM\ and gives the consistent gauge algebra.

The resulting form of the gauge transformation
is given by
$$
 \eqalign{
  & \delta (c) h_\mu^A
  = \partial_\mu c^A + {\partial W_{BC}(\Phi) \over \partial \Phi_A }
  h_\mu ^B c^C, \cr
  & \delta (c) \Phi_A = W_{BA} (\Phi) c^B \cr
 }
 \eqn\APHITR
$$
in terms of the functions $W_{AB}$ which satisfy the condition \JACPHI.
This gauge transformation reduces to the usual nonabelian one
when $W_{BA}(\Phi) = f_{BA}^C \Phi_C$.

Then the gauge algebra is given by
$$
 \eqalign{
  &[\delta (c_1), \delta (c_2)]h_\mu ^A \,= \delta (c_3)h_\mu ^A
             - c_1^Cc_2^D { \partial^2 W_{CD} \over
		\partial \Phi_A \partial \Phi_B } D_\mu \Phi _B, \cr
  &[\delta (c_1), \delta (c_2)]\Phi _A = \delta (c_3)\Phi _A, \cr
 }
 \eqn\FINGA
$$
where we have defined
$$
  D_\mu \Phi _A = \partial_\mu \Phi _A + W_{AB}(\Phi) h_\mu ^B.
 \eqn\DEFCOV
$$
We see the gauge transform of \DEFCOV:
$$
  \delta (c) (D_\mu \Phi _A) = (D_\mu \Phi _C)
		{ \partial W_{BA} \over \partial \Phi_C} c^B.
 \eqn\DFTRA
$$
This reveals that the object $D_\mu $ may be recognized as a covariant
differentiation, $D_\mu \Phi _A$ transforming as a coadjoint vector.
If the algebra is nondegenerate,
it turns out that the gauge algebra is open
as seen from \FINGA\ except for the case
${ \partial^2 W_{CD} \over \partial \Phi_A \partial \Phi_B } = 0 $,
that is, the case of nonabelian gauge transformation.

The commutator of two covariant differentiations
provides
$$
  [D_\mu , D_\nu ]\Phi _A = W_{AB}(\Phi)R_{\mu\nu}^B,
 \eqn\COMMD
$$
where the curvature $R_{\mu \nu}^A$ is defined by
$$
  R_{\mu \nu }^A = \partial_\mu h_\nu ^A - \partial_\nu h_\mu ^A
            + {\partial W_{BC} \over \partial \Phi_A} h_\mu ^B h_\nu ^C,
 \eqn\CURVAT
$$
and the covariant differentiation on $D_\mu \Phi _A$ has been
so determined
as the resultant expression $D_\nu (D_\mu \Phi _A)$ transforms
like a coadjoint field, that is,
$$
D_\mu D_\nu \Phi_A =  \partial_\mu D_\nu \Phi_A
	+ { \partial W_{AB} \over \partial \Phi_C }
	h_\mu{}^B D_\mu \Phi_C.
$$
Unfortunately, the curvature does not transform homogeneously
because the gauge algebra \FINGA\ is open:
$$
  \delta (c) R_{\mu \nu }^A = { \partial W_{BC} \over
		\partial \Phi_A} R_{\mu \nu }^B c^C
                + \left\{ (D_\mu \Phi _D) { \partial^2 W_{BC} \over
		  \partial \Phi_D \partial \Phi_A} h_\nu ^B c^C
                 - (\mu  \leftrightarrow \nu ) \right\},
 \eqn\TRACUR
$$
which seems troublesome for the construction of invariant actions.

%
\section{Jacobi Identity for Nonlinear Lie Algebra}

This section is devoted to a brief explanation
of an algebraic background for the `nonlinear' gauge transformation
defined in the previous section. The relevant object is
the Jacobi identity for nonlinear Lie algebras\rlap.
\refmark{\Sch, \Iza, \Dil}

We consider a vector space with a basis $\{ T_A \}$
which is the polynomial algebra with a commutative product.
On this algebra, we define
a bracket product
$$
  [T_A, T_B] = W_{AB}(T),
 \eqn\NLLALG
$$
as the second product structure,
where $W_{AB}(T)$ denotes a polynomial in $T_C$
which satisfies the antisymmetry property
$W_{AB} = -W_{BA}$:
$$
  W_{AB}(T) = W^{(0)}_{AB} + W^{(1)C}_{AB}T_C + W^{(2)CD}_{AB}T_CT_D
              + \cdots,
 \eqn\POLYN
$$
where $W^{(0)}_{AB}$, $W^{(1)C}_{AB}$, etc.~stand
for structure constants
in this algebra.
We assume the bracket product is
bilinear and antisymmetric with respect to exchange $A$ with $B$,
satisfies the Jacobi identity and
has the derivation nature
$$\eqalign{
& [T_A, T_B T_C] = [T_A, T_B]T_C + T_B[T_A, T_C] \cr
& [T_A T_B, T_C] =  T_A[T_B, T_C] + [T_A, T_C]T_B. \cr
}
$$

We emphasize that the multiplication of $T_C$ and $T_D$ is
defined in the sense of the polynomial
algebra (i.e. not as the matrix, for instance),
thus $T_CT_D = T_DT_C$.
Generally speaking, it is not necessary to
restrict consideration
to the case of polynomial algebra, which is commutative by definition.
However, it is enough to treat the commutative case
for the present purposes
because we make use of the structure functions $W_{AB}$ only in the form
$W_{AB}(\Phi)$ where $\Phi$ represents a set of scalar fields
$\Phi_C$ which do commute: $\Phi_C \Phi_D = \Phi_D \Phi_C$.

We note that the zeroth-order term $W^{(0)}_{AB}$ may
be recognized as a central element
in the algebra \NLLALG.
This definition includes
the ordinary Lie algebras as a special case when $W_{AB}(T)$
is linear in $T_A$, which makes the algebra \NLLALG\ deserve the name
of nonlinear Lie algebra.

The typical examples of nonlinear algebra are
W algebra\refmark{\Zam} and quantum group\rlap.\refmark{\Dri}
\foot{Strictly speaking, it is needed to extend $W_{AB}(T)$
to analytic functions of $T_A$ in the definition of nonlinear Lie
algebra.}
The representation theories of quadratic extension of
nonlinear $SU(2)$ algebra
and some interesting nonlinear algebras are
investigated by several authors\rlap.\refmark{\Roc}

The Jacobi identity with respect to the bracket product \NLLALG\
implies
$$
  { \partial W_{[AB} \over \partial T_D }
   W_{C]D} = 0,
 \eqn\JACT
$$
which is a generalization of the Jacobi identity
for Lie algebras.

The expression \FINGA\ can be regarded as a gauge algebra
based on the nonlinear Lie algebra \NLLALG\
in the sense that the functions $W_{AB}(\Phi)$ satisfy
the same constraint identity \JACPHI\ as
the Jacobi identity \JACT\ for $W_{AB}(T)$ in the algebra \NLLALG.
However, generally the field $\Phi_A$
is not a representation of
its algebra different from the coadjoint field of Lie
algebra.
The relation of $\Phi_A$ with the algebra
should be more investigated.

%
\section{An Invariant Action}

We do not have a straightforward way of constructing
invariant actions under the gauge transformation \APHITR\
owing to the inhomogeneous form of \TRACUR\
as noticed in section 2.1.
Hence we will instead seek for an action which provides
covariant equations of motion under those transformations,
and examine whether it is invariant under them.

We determine of equations of motion
as follows:
The gauge algebra \FINGA\ should be closed,
at least, on shell so as to originate from some Lagrangian symmetry.
Hence the choice
$$
  D_\mu \Phi_A = 0
 \eqn\FEOM
$$
seems an immediate guess for appropriate equations of motion,
which are indeed covariant \DFTRA\ under the transformation
\APHITR.

Since we have the vector, $h_\mu^A$, and the scalar, $\Phi_A$, fields
as our basic ones, as a first trial for Lagrangian
we consider
$$
  {\cal L}_0 = \chi ^{\mu \nu }h_\mu ^AD_\nu \Phi _A,
 \eqn\GUESS
$$
where $\chi ^{\mu \nu }$ denotes an invertible constant tensor
to be determined later.
However, $D_\nu \Phi _A$ itself contains the fields $h_\mu ^A$,
and hence the variation of ${\cal L}_0$ with respect to $h_\mu ^A$
does not lead to the covariant equations of motion \FEOM.

In order to overcome this issue,
we modify the Lagrangian into
$$
  {\cal L} = {\cal L}_0
            - {1\over 2}\chi ^{\mu \nu } W_{AB} h_\mu ^A h_\nu ^B,
 \eqn\TRUE
$$
which yields the desired equations of motion \FEOM\
provided $\chi ^{\mu \nu } = -\chi ^{\nu \mu }$.
Note that this Lagrangian can be rewritten as
$$
  {\cal L} = -{1\over 2}\chi ^{\mu \nu }[ \Phi _A R_{\mu \nu }^A
          + \left(W_{BC}- \Phi_A {\partial W_{BC} \over \partial
	  \Phi_A }\right) h_\mu ^Bh_\nu^C]
 \eqn\LAGCUR
$$
up to a total derivative.

If we require Poincar{\' e} invariance of the Lagrangian,
the antisymmetric tensor $\chi ^{\mu \nu }$ should be an invariant one
$$
  \chi ^{\mu \nu } = \epsilon ^{\mu \nu },
 \eqn\EPSILON
$$
where $\epsilon ^{\mu \nu }$ denotes
the Levi-Civita tensor in two dimensions.
At this point, we are inevitably led to considering
 the two-dimensional situation.

In summary, we have obtained a candidate action
$$
  S = \int d^2 x \, {\cal L}; \quad
  {\cal L} = -{1\over 2}\epsilon^{\mu \nu } \left[ \Phi _A R_{\mu \nu }^A
            + \biggl(W_{BC}- \Phi_A
           {\partial W_{BC} \over \partial \Phi_A }\biggr)
	  h_\mu ^Bh_\nu ^C \right],
 \eqn\LAGL
$$
which comes out to be diffeomorphism invariant.
In fact, the gauge transformation of ${\cal L}$ is
$$\eqalign{
  \delta (c){\cal L} & =
	\partial_\mu \left[ \epsilon^{\mu\nu}
	\biggl( W_{BC} - \Phi_A {\partial W_{BC} \over \partial \Phi_A }
	  \biggr) h_\nu^B c^C \right]
	+ \epsilon^{\mu\nu} \biggl[ - \partial_\mu W_{BC} h_\nu{}^B c^C \cr
	& \quad
	+ \partial_\mu \Phi_A {\partial W_{BC} \over \partial \Phi_A }
	h_\nu{}^B c^C
	+ ( W_{BD} {\partial W_{CE} \over \partial \Phi_B } + {1 \over 2}
	 {\partial W_{CD} \over \partial \Phi_B } W_{EB} )
	h_\mu{}^C h_\nu{}^D c^E \biggr], \cr
}%
$$
The relation \JACPHI\ enables us to show
$$
  \delta (c) {\cal L} =
	\partial_\mu \left[ \epsilon^{\mu\nu}
	\biggl( W_{BC} - \Phi_A {\partial W_{BC} \over \partial \Phi_A }
	  \biggr) h_\nu^B c^C \right],
 \eqn\LAGTR
$$
which confirms the invariance of our action \LAGL\ under
the `nonlinear' gauge transformation \APHITR.
The equations of motion which follow from \LAGL\ are given by
the following:
$$
  \epsilon ^{\mu \nu }D_\nu \Phi _A = 0,
  \quad \epsilon ^{\mu \nu }R_{\mu \nu }^A = 0,
 \eqn\EQMO
$$
which are indeed covariant
due to the transformation law \DFTRA, \TRACUR.

%
\section{BRS Formalism}

Since the action \LAGL\ has the local gauge symmetry
by the gauge transformation \APHITR,
there are some constraints.
In order to quantize the theory in such system,
it is necessary to fix gauge.
When we make gauge-fixing and quantize gauge theory,
use of BRS formalism is standard.

In this section,
we formulate BRS formalism of the `nonlinear' gauge theory.

We discuss how to construct
the BRS transformation $\brs$
corresponding to the classical symmetry \APHITR.
Notice that the generator algebra
for the gauge transformation \APHITR\ is open\rlap,\refmark{\Bat}
as can be seen from
the gauge algebra \FINGA.

We first try to regard the gauge-transformation
parameter $c^A$ as fermionic FP ghost,
$$
 \eqalign{
  \brssub h_\mu^A
   &= \partial_\mu c^A + U_{BC}^A(\Phi) h_\mu^B c^C, \cr
  \brssub \Phi_A &= W_{BA}(\Phi) c^B, \cr
 }
 \eqn\GBRS
$$
where $\brssub$ is a candidate for the BRS transformation.
In this level, $c^A$ is not a gauge parameter like
$c^A$ in the section 2.2,
but a fundamental field.
With the aid of the BRS nilpotency
condition $\brssub^2 \Phi _A = 0$,
we can uniquely
determine the transformation law for the ghost
whether $c^A$ depends on $\Phi_A$ or not,
which is different from the classical gauge transformation
in section 2.1;
$$
 \eqalign{
  \brssub c^A &= -{1 \over 2}
               {\partial W_{BC} \over \partial \Phi_A} c^B c^C, \cr
 }
 \eqn\GFPG
$$
which satisfies $\brssub^2 c^A = 0$.
However, the naive definition \APHITR\
does not satisfy the nilpotency $\brssub^2 =0$ on the field
$h_\mu ^A$ owing to the algebra non-closure:
$$
 \eqalign{
  \brssub^2 h_\mu^A &= -c^B c^C { \partial^2 W_{BC} \over \partial \Phi_A
		    \partial \Phi_D } D_\mu \Phi_D \cr
                 &= c^B c^C { \partial^2 W_{BC} \over \partial \Phi_A
		    \partial \Phi_D }
                      \epsilon_{\mu\nu} {\delta {\cal L}
                      \over \delta h_\nu^D}. \cr
 }
 \eqn\SQUAR
$$
This prevents us from performing simple-minded BRS gauge-fixing of
the Lagrangian \LAGL.
{}From the general discussion\refmark{\Bat},
it is sufficient that the BRS transformation is
off-shell symmetry on the quantum action $S_T$:
$$
\brs S_T = 0,
$$
and on-shell nilpotent on all fields
under quantum equations of motion
because Heisenberg operators always satisfy
quantum equations of motion.

The desired BRS transformation $\brs$ and gauge-fixed Lagrangian
${\cal L}_T$ can be obtained
as follows:
In view of the equation \SQUAR,
we introduce a gauge-fixing fermionic function $\Omega $
to provide
$$
  \brs = \brssub + \brso,
 \eqn\BRS
$$
where we define
$$
  \brso h_\mu^A = -i c^B c^C  { \partial^2 W_{BC} \over \partial \Phi_A
		    \partial \Phi_D }
                     \epsilon_{\mu\nu}
                    {\delta \Omega  \over \delta h_\nu^D},
 \eqn\BRSO
$$
and $\brso = 0$ on the other fields.
The gauge-fixed Lagrangian is given by
$$
  {\cal L}_T = {\cal L} - i(\brs - {1 \over 2}\brso)\Omega .
 \eqn\QUANL
$$

Then $\brs^2 h_\mu{}^A$ is calculated to be
$$\eqalign{
\brs^2 h_\mu{}^A & =
	c^B c^C { \partial W_{BC} \over \partial \Phi_A \partial \Phi_D }
	\epsilon_{\mu\nu} {\delta \over \delta h_\nu{}^D } \left[ {\cal L}
	- i (\brs - {1 \over 2}\brso)\Omega \right]  \cr
	& \quad + { \partial^2 \over \partial \Phi_A \partial \Phi_D }
	\left( i c^B c^C c^F {\partial W_{BC} \over \partial \Phi_E }
	W_{EF} \right) {\delta \Omega \over \delta h_\mu{}^D }, \cr
}
\eqn\BRSONH
$$
by using
$$\eqalign{
{\delta \over \delta h_\mu{}^D }
\left[ ( \brssub + {1 \over 2} \brso ) \Omega \right]
& = ( \brssub + \brso ) \Omega + {\delta \Omega \over \delta h_\mu{}^G }
{\partial W_{DF} \over \partial \Phi_G } c^F. \cr
}
$$
Since the Jacobi identity \JACPHI\ provides the following identity
$$
c^B c^C c^F {\partial W_{BC} \over \partial \Phi_E } W_{EF} = 0,
$$
it turns out that $\brs^2 h_\mu{}^A$ vanishes
under quantum equations of motion,
that is, $\brs$ is on-shell nilpotent,
and \BRS\ yields a symmetry of the Lagrangian \QUANL,
as it should be:
$$\eqalign{
  \brs {\cal L}_T & =
	\partial_\mu \left[ \epsilon^{\mu\nu}
	\biggl( W_{BC} - \Phi_A {\partial W_{BC} \over \partial \Phi_A }
	  \biggr) h_\nu^B c^C \right] \cr
        & \quad + \partial_\mu \biggl(i \Phi_A c^B c^C
	{ \partial^2 W_{BC} \over \partial \Phi_A \partial \Phi_D }
     {\delta \Omega  \over \delta h_\mu^D} \biggr). \cr
}
 \eqn\BRSD
$$

%
\section{Quantization on the Cylinder}

By means of the BRS formalism given in the previous section,
we quantize the system
on the cylindrical spacetime
$S^1\!\times{\mib R}^1$
in the temporal gauge.

The gauge-fixing function for the temporal gauge is
$$
  \Omega = {\bar c}_{\,A}h_0{}^A,
 \eqn\TGFF
$$
where we introduce FP antighosts ${\bar c}_{\,A}$
and NL fields $b_A$ which satisfy the BRS transformation law
$$
  \brs {\bar c}_{\,A} = ib_A.
 \eqn\BRSAG
$$
The gauge-fixed Lagrangian \QUANL\ now reads
$$
 \eqalign{
  {\cal L}_T &= {\cal L} + b_A h_0{}^A
		+ i{\bar c}_A ( {\dot c}^A
		+ { \partial W_{BC} \over \partial \Phi_A }h_0{}^B c^C ), \cr
 }
$$
where the dot denotes a derivative with respect to the time $x^0$.
This Lagrangian reduces to the form of a free theory
$$
  {\cal L}^\prime_T = \Phi_A {\dot h}_1{}^A
              + b'_A h_0{}^A + i{\bar c}_A{\dot c}^A
 \eqn\FREE
$$
through field redefinition
$$
 \eqalign{
  b'_A &\equiv b_A + \partial_1 \Phi_A
               - W_{AB} h_1{}^B
		+ i {\bar c}_B {\partial W_{AC} \over \partial \Phi_B }
		c^C, \cr
 }
$$
where we have discarded total derivatives
w.r.t.~$x^1$ in the Lagrangian \FREE.

We can quantize explicitly \FREE\ on $S^1\!\times{\mib R}^1$
and derive mode expansion as follows:
$$\eqalign{
  h_1{}^A (x) & = { 1 \over \sqrt{2 \pi} } \sum^\infty_{n = -\infty}
	h_n{}^A e^{- i n x^1}, \cr
  \Phi_A (x) & = { 1 \over \sqrt{2 \pi} } \sum^\infty_{n = -\infty}
	\Phi_{An} e^{- i n x^1}, \cr
  c^A (x) & = { 1 \over \sqrt{2 \pi} } \sum^\infty_{n = -\infty}
	c_n^A e^{- i n x^1}, \cr
  {\bar c}_A (x) & = { 1 \over \sqrt{2 \pi} } \sum^\infty_{n = -\infty}
	{\bar c}_{An} e^{- i n x^1}. \cr
}
$$
{}From the canonical commutation relations of the fundamental fields,
we get the commutation relations of mode variables:
$$\eqalign{
& [ \Phi_{An}, h_m^B ] = - i \delta^B_A \delta_{n+m,0}, \cr
& \{ {\bar c}_{An}, c^B_m \} = \delta^B_A \delta_{n+m,0}. \cr
}
\eqn\MODECOM
$$
As usual,
it is assumed that $\Phi_{An}$, $h_n^A$,
${\bar c}_{An}$ and $c^A_n$ for $n \geq 1 $ annihilate
the Fock vacuum.
Then nilpotency of the BRS charge in
quantum level is necessary for the theory
to be unitary and to be well-defined.

The symmetry \BRS\ yields a conserved charge
$$
  Q = \int_{S^1} dx^1 J_{\hbox {\sc B}}^{\,0},
 \eqn\BRSCH
$$
where its Noether current
$J^\mu_{\hbox {\sc B}}$ is defined by
$$\eqalign{
  J^\mu_{\hbox {\sc B}} & =
	\brs h_\nu{}^A {\partial {\cal L}_T \over \partial
	(\partial_\mu h_\nu{}^A)}
	+ \brs \Phi_A {\partial {\cal L}_T \over \partial
	(\partial_\mu \Phi_A)}
	+ \brs c^A {\partial {\cal L}_T \over \partial
	(\partial_\mu c^A)}
	+ \brs {\bar c}_A {\partial {\cal L}_T \over \partial
	(\partial_\mu {\bar c}_A)}
	+ \brs b_A {\partial {\cal L}_T \over \partial
	(\partial_\mu b_A)} \cr
	& \quad -\epsilon^{\mu\nu}
        \biggl( W_{BC} - \Phi_A {\partial W_{BC} \over \partial \Phi_A }
          \biggr) h_\nu^B c^C
	- i \Phi_A c^B c^C  { \partial^2 W_{BC} \over \partial \Phi_A
                    \partial \Phi_D }
     {\delta \Omega  \over \delta h_\mu^D}. \cr
}
$$
The zeroth component of its Noether current
$J_{\hbox {\sc B}}$ is
given by
$$
 \eqalign{
  J_{\hbox {\sc B}}^{\,0} &=
  \Phi_A \partial_1 c^A + J; \cr
  J &= {1 \over 2} \{ W_{BC}, h_1{}^B \} c^C
	- {i \over 2 } { \partial W_{BC} \over \partial \Phi_A }
	c^B {\bar c}_A c^C. \cr
 }
 \eqn\BRSCU
$$
by using \TGFF.
Note that the operator ordering offers a matter for consideration
in the definition of $J$ written by mode variables.
The ordering should
have been so chosen as the BRS charge \BRSCH\ to be hermitian
like \BRSCU.

This naive definition of the charge
suffers from divergence difficulty,
which can be removed
by normal-ordering prescription of mode variables
$Q_{\hbox {\sc B}}\! =\ :\!\!Q\!\!:$\ .
$$\eqalign{
Q_{\hbox {\sc B}} & = i \sum_{n=-\infty}^{\infty} n \lnormal\Phi_{An}
	c_{-n}^A \rnormal \cr
	& \quad + {1 \over 2} \sum_k
	\sum_{n_1=-\infty}^{\infty} \scdots
	\sum_{n_k=-\infty}^{\infty} \sum_{m=-\infty}^{\infty}
	 W_{BC}^{(k)A_1\cdots A_k}\lnormal \{\Phi_{A_1n_1}\cdots\Phi_{A_kn_k}
	h_{1m}^B \} c^C_{-\sum_k n_i - m}\rnormal \cr
	& \quad - {i \over 2} \sum_k
	\sum_{n_1=-\infty}^{\infty} \scdots
	\sum_{n_k=-\infty}^{\infty}
	\sum_{\scriptstyle m=-\infty\atop\scriptstyle l=-\infty}^{\infty}
	W_{BC}^{(k)A_1\cdots A_k} \lnormal \Phi_{A_1n_1}\cdots
	\Phi_{A_{k-1}n_{k-1}} c^B_{m}{\bar c}_{Al}
	c^C_{-\sum_{k-1} n_i - m - l}\rnormal, \cr
}
$$
The canonical commutation relation
\MODECOM\ establishes the nilpotency
$Q_{\hbox {\sc B}}^2 = 0$
of the BRS operator straightforwardly.
This enables us to define the quantum theory
in terms of the BRS cohomology\rlap.
\refmark{\Kug}

%
\chapter{$R^2$ gravity with dynamical torsion}
%


In this chapter, we consider $R^2$ gravity
with dynamical torsion in two spacetime dimensions
\refmark{\Kat} as the typical example of
the nonlinear gauge theory\rlap.\refmark{\Iza}

%
\section{Pure $R^2$ Gravity}

We first make a brief exposition of $R^2$ gravity
with dynamical torsion in two spacetime dimensions\rlap.
\refmark{\Kat}

When the spin connection $\omega_\mu{}^{ab}$
and the zweibein $e_\mu{}^a$
are independent variables, the action, invariant under
local Lorentz and general coordinate transformations
$$
 \eqalign{
  \gct \omega_\mu &= \partial_\mu \tau
                     -v^\lambda \partial_\lambda \omega_\mu
                     -(\partial_\mu v^\lambda) \omega_\lambda, \cr
  \gct e_\mu{}^a &= -\tau \epsilon^{ab}e_{\mu b}
	            -v^\lambda \partial_\lambda e_\mu{}^a
                    -(\partial_\mu v^\lambda) e_\lambda{}^a, \cr
 }
 \eqn\DIFFE
$$
generically
\foot{We do not consider higher-derivative theories\rlap.
\refmark{\Yon}}
takes the following form:
$$
  S = \int d^2x
       \ e({1 \over 16 \alpha}R_{\mu\nu}{}^{ab}R^{\mu\nu}{}_{ab}
        - {1\over 8 \beta}T_{\mu\nu}{}^{a}T^{\mu\nu}{}_{a} - \gamma),
 \eqn\GRAVA
$$
where $\tau$ and $v^\lambda$ are local-Lorentz and general coordinate
transformation
parameters ,and $\alpha$, $\beta$, $\gamma$ are constant parameters,
and
$$
 \eqalign{
  \omega_\mu\epsilon^{ab} &\equiv \omega_\mu{}^{ab},
   \quad e \equiv \det (e_\mu{}^a), \cr
  R_{\mu\nu}{}^{ab} &\equiv \partial_\mu \omega_\nu{}^{ab}
	             -\partial_\nu \omega_\mu{}^{ab}, \cr
  T_{\mu\nu}{}^a &\equiv \partial_\mu e_\nu{}^a
	          +\omega_\mu{}^{ab}e_{\nu b}
                   -(\mu \leftrightarrow \nu). \cr
 }
$$
Our convention is that
the Levi-Civita antisymmetric tensor $\epsilon$ satisfies
$\epsilon_{01}=\epsilon^{10}=1$, and
the flat metric is given by $\eta = {\rm diag}(+1, -1)$.

The Lagrangian in the action \GRAVA\ can be rewritten as
$$
  {\cal L}_G = {1 \over 4e\alpha} (F_{01})^2
             + {1 \over 4e\beta}T_{01}{}^aT_{01a} - e\gamma,
 \eqn\GRAVL
$$
where we have defined
$$
  F_{\mu\nu} \equiv \partial_\mu \omega_\nu - \partial_\nu \omega_\mu.
$$
Various aspects of this theory
were extensively studied in Ref.[\Kat].

%
\section{Yang-Mills-like Formulation}

In order to interpret the theory defined by \GRAVA\ as
a nonlinear gauge theory,
we consider
a first-order Lagrangian
$$
  {\cal L}_C = \varphi F_{01} - e \alpha \varphi^2
             + \phi_a T_{01}{}^a - e \beta \phi_a \phi^a - e \gamma
 \eqn\TOPOL
$$
by introducing auxiliary fields $\varphi$ and $\phi_a$.
This is equivalent to \GRAVL\
in the region $e \ne 0$.
Namely, it coincides with the Lagrangian \GRAVL\
when the auxiliary fields
$\varphi$ and $\phi_a$ are integrated out.
The Lagrangian \TOPOL\ also has
symmetry under local Lorentz and general coordinate
transformations \DIFFE\ with
$$
 \eqalign{
  \gct \varphi &= -v^\lambda \partial_\lambda \varphi, \cr
  \gct \phi_a &= -\tau \epsilon_{ab}\phi^b
                 -v^\lambda \partial_\lambda \phi_a. \cr
 }
 \eqn\DIFFA
$$
Its equations of motion are as follows:
$$
 \eqalign{
  &\partial_\mu \varphi + \phi_a \epsilon^{ab} e_{\mu b} = 0, \cr
  &\partial_\mu \phi_a + \omega_\mu \epsilon_{ab} \phi^b
   - \epsilon_{ab} e_\mu{}^b
    (\alpha \varphi^2 +\beta \phi_c\phi^c + \gamma) = 0, \cr
  &2e\alpha \varphi - F_{01} = 0, \quad
    2e\beta \phi^a - T_{01}{}^a = 0. \cr
 }
 \eqn\EQOM
$$

We note that the Lagrangian \TOPOL\
in a weak-coupling limit $\alpha = \beta = \gamma = 0$
is none other than
that of topological ISO(1,\thinspace 1) gauge theory\rlap,
\refmark{\Fuk}
which is known to be equivalent to
the gravity theory\refmark{\Jac}
$$
  I = \int d^2x \sqrt{-g} \sigma R.
 \eqn\JACKIW
$$
This observation suggests that the full Lagrangian \TOPOL\
itself might have some symmetry of the Yang-Mills type
which is related to the ISO(1,\thinspace 1) gauge transformation.

In order to find such symmetry,
we use a trivial gauge symmetry proportional to
the equations of motion \EQOM,
which Lagrangian system always has.
We modify the local-Lorentz and general coordinate
transformation \DIFFE\ and \DIFFA\ by the trivial gauge symmetry
$$
 \eqalign{
  \iso \omega_\mu &= \gct \omega_\mu +\epsilon_{\mu\lambda}v^\lambda
   [2 \alpha e \varphi - F_{01}], \cr
  \iso e_\mu{}^a &= \gct e_\mu{}^a + \epsilon_{\mu\lambda}v^\lambda
   [2 \beta e \phi^a - T_{01}{}^a], \cr
  \iso \varphi &= \gct \varphi + v^\lambda [\partial_\lambda \varphi
                         + \phi_a \epsilon^{ab} e_{\lambda b}], \cr
  \iso \phi_a &= \gct \phi_a + v^\lambda
   [\partial_\lambda \phi_a + \omega_\lambda \epsilon _{ab} \phi^b
    - \epsilon_{ab} e_\lambda{}^b
     (\alpha \varphi^2 +\beta \phi_c\phi^c + \gamma)]. \cr
 }
 \eqn\BETW
$$

If we set
$$
  t = \tau -v^\lambda \omega_\lambda,
  \quad c^a = -v^\lambda e_\lambda{}^a,
$$
we can explicitly write down
\foot{Note added: Quite recently, this symmetry has been also
obtained
in Ref.[\Str] by means of Hamiltonian analysis of the action \GRAVA.}
the symmetry $\iso$:
$$
 \eqalign{
  \iso \omega_\mu &= \partial_\mu t
                    + 2\alpha \epsilon_{bc} c^b e_\mu{}^c \varphi, \cr
  \iso e_\mu{}^a &= - t \epsilon^{ab} e_{\mu b}
                 + \partial_\mu c^a + \omega_\mu \epsilon^{ab} c_b
                  + 2 \beta \epsilon_{bc} c^b e_\mu{}^c \phi^a, \cr
  \iso \varphi &= \epsilon^{ab} c_a \phi_b, \cr
  \iso \phi_a &= - t \epsilon_{ab} \phi^b
                + \epsilon_{ab} c^b(\alpha \varphi^2
                + \beta \phi_c \phi^c + \gamma), \cr
 }
 \eqn\MISO
$$
which reduces to the ISO(1,\thinspace 1) gauge transformation
\foot{Instead of the Poincar{\' e} group ISO(1,\thinspace 1),
we can deal with the de Sitter group SO(2,\thinspace 1)
by means of a constant shift in the field $\varphi$
and an appropriate redefinition
of the parameter $\gamma$.}
when $\alpha = \beta = \gamma = 0$.
It is straightforward to see
$$
  \iso {\cal L}_C
  = \partial_\mu [\epsilon^{\mu\nu}
   e_\nu{}^a \epsilon_{ab} c^b ( \alpha \varphi^2
    + \beta \phi_c \phi^c -\gamma )].
 \eqn\ISOD
$$
The gauge transformations \DIFFE, \DIFFA, and \MISO\ are reducible.

In the following sections,
we confirm \MISO\ is a special case of our nonlinear
gauge transformation introduced in the previous chapter.

%
\section{Quadratically Nonlinear Poincar{\' e} Algebra}
%


The general-covariance property of our action \LAGL\
suggests that we might utilize it to construct
two-dimensional gravitation theory.
It seems a natural choice to adopt the
Poincar{\' e} algebra in two dimensions
as a base algebra on which the construction will be performed.
Since the diffeomorphism invariance is guaranteed by the definition
of \LAGL, our requirement is that the theory should be local-Lorentz
covariant. Hence we consider quadratically nonlinear
extension of the Poincar{\' e} algebra which preserves
the Lorentz structure of the genuine Poincar{\' e} algebra:
$$
 \eqalign{
  &[ \,J, J\, ] = 0, \quad
   [ \,J, P_a ] = \epsilon_a{}^b P_b, \cr
  &[ P_a, P_b ] = - \epsilon_{ab} (\alpha J J
		   + \beta \eta^{cd} P_c P_d
		    + \gamma I), \cr
 }
 \eqn\MODISO
$$
where $\alpha $, $\beta $, $\gamma $ are constants,
$\eta ^{cd}$ is the two-dimensional flat metric,
and $I$ is a central element added to the Poincar{\' e} algebra\rlap.
\foot{The right-hand side of $[P_a, P_b]$ can contain a term
proportional to $\epsilon _{ab}J$, which results in nonlinear (anti-)de Sitter
algebra. We omit that term just for simplicity.}

We set $\{ T_A \} = \{ P_a, J \}$, that is, $T_0 = P_0$, $T_1 = P_1$,
and $T_2 = J$.
Then the structure constants defined
for the above algebra are given by
$$
 \eqalign{
  W^{(1)a}_{CD} &= \epsilon^a{}_b( \delta_C{}^b \delta_D{}^2
	     - \delta_C{}^2 \delta_D{}^b ),
   \quad W^{(1)2}_{CD} = 0, \cr
  W^{(2)22}_{CD} &= - \alpha (\delta_C{}^k \delta_D{}^l
		\epsilon_{kl}), \quad
   W^{(2)ab}_{CD} = - \beta \eta^{ab} (\delta_C{}^k \delta_D{}^l
		\epsilon_{kl}),
    \quad W^{(2)2a}_{CD} = W^{(2)a2}_{CD} = 0, \cr
  W^{(0)}_{CD} &= - \gamma (\delta_C{}^k \delta_D{}^l \epsilon_{kl}), \cr
  W^{(i)} & = 0 \quad {\rm for }\ i \geq 3, \cr
}
$$
where the notation of \POLYN\ is used.

We also set gauge fields $h_\mu ^A = (e_\mu ^a, \omega _\mu )$ and
scalar fields $\Phi _A = (\phi _a, \varphi)$ to obtain
\TOPOL:
$$
  {\cal L}_1 = - {1\over 2}\epsilon ^{\mu \nu } \varphi F_{\mu \nu }
               - {1\over 2}\epsilon ^{\mu \nu } \phi_a T_{\mu \nu }{}^a
               - e \alpha \varphi^2 - e \beta \phi_a \phi^a - e \gamma
$$
as the Lagrangian
\LAGL.
The gauge transformation law
\APHITR\
now reads
\MISO:
$$
 \eqalign{
  \iso \omega_\mu &= \partial_\mu t
                    + 2\alpha \epsilon_{bc} c^b e_\mu{}^c \varphi, \cr
  \iso e_\mu{}^a &= - t \epsilon^{ab} e_{\mu b}
                 + \partial_\mu c^a + \omega_\mu \epsilon^{ab} c_b
                  + 2 \beta \epsilon_{bc} c^b e_\mu{}^c \phi^a, \cr
  \iso \varphi &= \epsilon^{ab} c_a \phi_b, \cr
  \iso \phi_a &= - t \epsilon_{ab} \phi^b
                + \epsilon_{ab} c^b(\alpha \varphi^2
                + \beta \phi_c \phi^c + \gamma), \cr
 }
$$
where we have put $c^A = (c^a, t)$.

This theory completely coincides with the Yang-Mills-like
formulation of $R^2$ gravity with dynamical torsion
we proposed the previous section.
Note that the gravitation theory \GRAVA\ is obtained
if one integrates out the scalar fields $\varphi$ and $\phi_a$
in the Lagrangian \TOPOL\ under the condition $e \neq 0$.
In short, two-dimensional $R^2$ gravity with dynamical torsion
has been shown to be a gauge theory based on quadratically
nonlinear Poincar{\' e} algebra.

%
\chapter{Generic Form of `Dilaton' Gravity}

In this chapter, we consider generic form of 'dilaton' gravity
as the nonlinear gauge theory\rlap.\refmark{\Dil}

%
\section{The Action}

The general expression for the action of two-dimensional metric
${\bar g}_{\mu \nu}$ coupled to a scalar field $\bar{\varphi}$
(without higher-derivative terms)
is given by
$$
  S = \int \! d^2 x \, \sqrt{- \bar g} \,( {1 \over 2} \bar g^{\mu\nu}
       \partial_\mu \bar{\varphi}  \partial_\nu \bar{\varphi}
	+ {1 \over 2}\, {\cal U}(\bar{\varphi} ) \bar R
         - {\cal V}(\bar{\varphi})).
 \eqn\GDGR
$$
We may
replace ${\cal U}({\bar \varphi})$ by a linear function
\refmark{\Rus}
$$
  S = \int \! d^2 x \, \sqrt{-g}
    \,({\kappa \over 2} g^{\mu\nu} \partial_\mu \varphi
     \partial_\nu \varphi + {1 \over 2}\varphi
      R - {\cal W}(\varphi) )
 \eqn\MIDDLE
$$
through a field redefinition
\foot{This field transformation is well-defined only locally
in the field space of ${\bar \varphi}$
where $(\partial {\cal U}/\partial {\bar \varphi}) \neq 0$.
In particular, it is inapplicable to the case of
${\cal U}({\bar \varphi}) = {\rm constant}$,
where the field ${\bar \varphi}$
is regarded as a scalar matter rather than the `dilaton' field.}
$$
 \eqalign{
  & \varphi = {\cal U}(\bar{\varphi}),
   \quad g_{\mu\nu} = e^{\rho} \bar{g}_{\mu\nu},
    \quad {\cal W}(\varphi) = e^{\rho} {\cal V}(\bar{\varphi}); \cr
  & \rho( \bar{\varphi}) \equiv -{\kappa}\, {\cal U}(\bar{\varphi})
    + \int \!
   d \bar{\varphi} \,
    \biggl({\partial {\cal U}
     \over \partial \bar{\varphi}} \biggr)^{-1}, \cr
 }
 \eqn\SECACT
$$
where $\kappa$ is an arbitrary constant.
In particular, we can set $\kappa = 0$ to obtain
$$
  S =  \int \! d^2x \,{\cal L}_D; \quad
  {\cal L}_D = \sqrt{- g} \,({1 \over 2}\varphi R - {\cal W}(\varphi)).
 \eqn\DACTION
$$
The action \DACTION\ is of course invariant under
the following general coordinate transformation:
$$
 \eqalign{
  \gct g_{\mu\nu} &= -v^\lambda \partial_\lambda g_{\mu\nu}
                     -(\partial_\mu v^\lambda) g_{\lambda\nu}
                     -(\partial_\nu v^\lambda) g_{\mu\lambda}, \cr
  \gct \varphi &= -v^\lambda \partial_\lambda \varphi. \cr
 }
 \eqn\DIFFE
$$

In the following section, we investigate
a gauge-theoretical origin of the `dilaton' field $\varphi$
in the action \DACTION, which might be regarded as a ``pure" gravity
in two dimensions.

%
\section{General Nonlinear Poincar{\' e} Algebra}
%

It seems a natural choice to adopt the
Poincar{\' e} algebra in two dimensions
as a base algebra on which the construction will be performed.
Since the diffeomorphism invariance is guaranteed by the definition
of \LAGL, our requirement is that the theory should be local-Lorentz
covariant. Hence we consider nonlinear extension
\foot{The following form of nonlinear Poincar{\' e} algebra
is not the most general one that preserves the Lorentz structure.
It is chosen so as to realize
the torsion-free condition as an equation of motion
in the resultant `nonlinear' gauge theory.}
of the Poincar{\' e} algebra which preserves
the Lorentz structure of the genuine Poincar{\' e} algebra:
$$
 \eqalign{
  &[ \,J, J\, ] = 0, \quad
   [ \,J, P_a ] = \epsilon_{ab} \eta^{bc} P_c, \cr
  &[ P_a, P_b ] = - \epsilon_{ab} {\cal W} (J), \cr
 }
 \eqn\MODISO
$$
where $\eta^{cd}$ is the two-dimensional Minkowski metric.
Note that the choice ${\cal W}(J) = 0$ corresponds to the original
Poincar{\' e} algebra.

We set $\{ T_A \} = \{ P_a, J \}$, that is, $T_0 = P_0$, $T_1 = P_1$,
and $T_2 = J$.
Then the structure functions defined in \NLLALG\
for the above algebra are given by
$$
 \eqalign{
  W_{2a} &= - W_{a2} = \epsilon_{ab} \eta^{bc} P_c,
   \quad W_{22} = 0, \cr
  W_{ab} & = - \epsilon_{ab} {\cal W}(J). \cr
 }
 \eqn\HATENA
$$

We also set the vector field $h_\mu^A = (e_\mu{}^a, \omega_\mu)$ and
the scalar field $\Phi_A = (\phi_a, \varphi)$ to obtain
$$
  {\cal L}_S = - {1\over 2}\epsilon ^{\mu \nu } \varphi F_{\mu \nu }
               - {1\over 2}\epsilon ^{\mu \nu } \phi_a T_{\mu \nu }{}^a
               - e {\cal W} (\varphi)
 \eqn\DTOPOL
$$
as the Lagrangian \LAGL,
where we have defined
$$
 \eqalign{
  F_{\mu\nu} &\equiv \partial_\mu \omega_\nu - \partial_\nu \omega_\mu,
   \cr
  T_{\mu\nu}{}^a &\equiv \partial_\mu e_\nu{}^a
	          +\omega_\mu\epsilon ^{ab}e_{\nu b}
                   -(\mu \leftrightarrow \nu), \cr
  e &\equiv {\rm det} (e_\mu{}^a). \cr
 }
 \eqn\DEFDEF
$$
The gauge transformation law \APHITR\
now reads
$$
 \eqalign{
  \iso \omega_\mu &= \partial_\mu t
                    + \epsilon_{bc} c^b e_\mu{}^c
               {\partial {\cal W}(\varphi) \over \partial \varphi}, \cr
  \iso e_\mu{}^a &= - t \epsilon^{ab} e_{\mu b}
                 + \partial_\mu c^a + \omega_\mu \epsilon^{ab} c_b, \cr
  \iso \varphi &= \epsilon^{ab} c_a \phi_b, \cr
  \iso \phi_a &= - t \epsilon_{ab} \phi^b
                + \epsilon_{ab} c^b {\cal W}(\varphi), \cr
 }
 \eqn\DMISO
$$
where we have put $c^A = (c^a, t)$.

We can actually confirm
$$
  \iso {\cal L}_S
   = -\partial_\mu [\epsilon^{\mu\nu}
   e_\nu{}^a \epsilon_{ab} c^b
   \biggl( {\cal W}
    - \varphi {\partial {\cal W} \over \partial \varphi} \biggr)],
 \eqn\ISOD
$$
which corresponds to \LAGTR\ in the previous section.
The equations of motion which follow from \DTOPOL\ are given by
$$
 \eqalign{
  &\partial_\mu \varphi + \phi_a \epsilon^{ab} e_{\mu b} = 0, \cr
  &\partial_\mu \phi_a + \omega_\mu \epsilon_{ab} \phi^b
   - \epsilon_{ab} e_\mu{}^b {\cal W} = 0, \cr
  &{1 \over 2} \epsilon^{\mu \nu} F_{\mu \nu}
    + e{{\partial {\cal W} \over \partial \varphi}} = 0, \quad
    {1 \over 2} \epsilon^{\mu \nu} T_{\mu \nu}{}^a = 0, \cr
 }
 \eqn\EQOM
$$
which of course corresponds to \EQMO.

Note that the gravitation theory \DACTION\ is obtained
if one puts $e_\mu{}^a e_\nu{}^b \eta_{ab} = g_{\mu\nu}$ and
integrates out the fields $\phi_a$ and $\omega_\mu$
in the Lagrangian \DTOPOL\ under the condition $e \neq 0$.
The reformulation \DTOPOL\ reveals that the theory \DACTION\ possesses
hidden gauge symmetry \DMISO\ of the Yang-Mills type.
This gauge symmetry includes diffeomorphism, as is the case
for the Yang-Mills-like formulation
of $R^2$ gravity with dynamical torsion in the
previous chapter.

%
\chapter{`Nonlinear'  Supergauge Theory}

In this chapter,
we extend the previous consideration to the
a `nonlinear' supergauge theory
i.~e.~gauge theory based on nonlinear extension of Lie superalgebras,
and apply this gauge theory to two-dimensional dilaton supergravity.

%
\section{`Nonlinear' Gauge Transformation}

We introduce a supermultiplet of gauge fields
$ ( h_\mu^A, \xi_\mu{}^\alpha)$,
where $ h_\mu^A $ is a bosonic field and
$\xi_\mu{}^\alpha$ a fermionic one.
$A$ denotes an internal index of the even generators
and $\alpha$ the odd generators.
As is the case for
`nonlinear' gauge theory
exposed in Ref.[\Sch, \Iza, \Dil],
we need a `coadjoint' bosonic scalar field $\phi_A$ and
a fermionic one $\psi_\alpha$.

According to the same discussion as in Ref.~[\Dil],
we determine the gauge transformation of gauge fields
$ h_\mu^A$ and $ \xi_\mu{}^\alpha $, auxiliary fields
$ \phi_A$ and $\psi_\alpha$.

By the same discussion as the bosonic case in the chapter 2,
the gauge transformation
of $h_\mu{}^A $, $\xi_\mu{}^\alpha $,
$\phi_A$ and $\psi_\alpha$ is determined.
The resulting form of the gauge transformation $\iso(c, \tau)$
is given by
$$\eqalign{
& \iso(c, \tau) h_\mu{}^A = \partial_\mu c^A
        - c^B \left( { \partial W_{BC} \over \partial \phi_A } h_\mu{}^C
        + i { \partial U_{B \gamma} \over \partial \phi_A } \xi_\mu{}^\gamma
        \right)
        - i \tau^\beta \left( { \partial U_{C\beta} \over \partial \phi_A}
        h_\mu{}^C + { \partial V_{\beta\gamma} \over \partial \phi_A }
        \xi_\mu{}^\gamma \right), \cr
& \iso(c, \tau) \xi_\mu{}^\alpha = \partial_\mu \tau^\alpha
        + c^B \left( - i { \partial W_{BC} \over \partial \psi_\alpha }
        h_\mu{}^C
        - { \partial U_{B \gamma} \over \partial \psi_\alpha}
        \xi_\mu{}^\gamma \right)
        + \tau^\beta \left( { \partial U_{C\beta} \over \partial \psi_\alpha}
        h_\mu{}^C + { \partial V_{\beta\gamma} \over \partial \psi_\alpha }
        \xi_\mu{}^\gamma \right), \cr
& \iso(c, \tau) \phi_A = c^B W_{BA} + i \tau^{\beta} U_{A \beta}, \cr
& \iso(c, \tau) \psi_\alpha
	= c^B U_{B\alpha} + \tau^{\beta} V_{\beta\alpha}, \cr
}\eqn\SUAPHITR$$
where
$c^A, \tau^\alpha$ are gauge parameters and
$W_{AB}( \phi, \psi )$, $U_{A \beta}( \phi, \psi )$ and
$V_{\alpha\beta}( \phi, \psi )$
are smooth functions of the fields
$\phi_A$ and $\psi_\alpha $ to be determined below.
Since the transformation includes the one
in the case of the usual Lie superalgebra,
we assume that
$W_{AB}( \phi, \psi ) = - W_{BA}( \phi, \psi )$
and
$V_{\alpha\beta}( \phi, \psi ) = V_{\beta\alpha}( \phi, \psi )$.
Similar to \JACPHI,
$W_{AB}$, $U_{A\beta}$ and $V_{\alpha\beta}$
should satisfy the condition
$$\eqalign{
& {\partial W_{AB} \over \partial \phi_D } W_{CD}
+ U_{A\delta} {\partial W_{BC} \over \partial \psi_\delta }
+ {\rm Antisym}(A, B, C) = 0, \cr
& {\partial W_{AB} \over \partial \phi_D } U_{D\gamma}
+ {\partial U_{B\gamma} \over \partial \phi_D } W_{DA}
- {\partial U_{A\gamma} \over \partial \phi_D } W_{DB}
+ U_{B\delta} {\partial U_{A\gamma} \over \partial \psi_\delta }
- U_{A\delta} {\partial U_{B\gamma} \over \partial \psi_\delta }
- i V_{\gamma\delta} {\partial W_{AB} \over \partial \psi_\delta }
= 0, \cr
& {\partial V_{\beta\gamma} \over \partial \phi_D } W_{DA}
+ i U_{D\gamma} {\partial U_{A\beta} \over \partial \phi_D }
+ i U_{D\beta} {\partial U_{A\gamma} \over \partial \phi_D }
+ V_{\beta\delta} {\partial U_{A\gamma} \over \partial \psi_\delta }
+ V_{\gamma\delta} {\partial U_{A\beta} \over \partial \psi_\delta }
+ {\partial V_{\beta\gamma} \over \partial \psi_\delta } U_{A\delta}
= 0, \cr
& i { \partial V_{\alpha\beta} \over \partial \phi_D } U_{D\gamma}
+ V_{\gamma\delta} {\partial V_{\alpha\beta} \over \partial \psi_\delta }
+ {\rm Sym}(\alpha, \beta, \gamma)
= 0, \cr
}\eqn\SUJACPHI
$$
where ${\rm Antisym}(A, B, C)$ denotes antisymmetrization
in the indices $A$, $B$ and $C$, and
\hfil \break
${\rm Sym}(\alpha, \beta, \gamma)$ symmetrization
in the indices $\alpha$, $\beta$ and $\gamma$.
The expression \SUJACPHI\
can be derived from the Jacobi identity for a nonlinear Lie superalgebra
with structure functions $W_{AB}$, $U_{A \beta}$ and $V_{\alpha\beta}$.

The gauge algebra is given by
$$\eqalign{
[ \iso( c_1, \tau_1 ), \iso(c_2, \tau_2 ) ] h_\mu{}^A
        & = \iso(c_3, \tau_3) h_\mu{}^A
        - c_1^B c_2^C \left( D_\mu \phi_D
        {\partial^2 W_{BC} \over \partial \phi_A \partial \phi_D}
        + D_\mu \psi_\delta
        {\partial^2 W_{BC} \over \partial \phi_A \partial \psi_\delta}
        \right) \cr
        & \quad + i ( \tau_2^B c_1^\gamma - \tau_1^\gamma c_2^B )
        \left( D_\mu \phi_D {\partial^2 U_{B\gamma} \over
        \partial \phi_A \partial \phi_D}
        + D_\mu \psi_\delta
        {\partial^2 U_{B\gamma} \over \partial \phi_A \partial
        \psi_\delta} \right) \cr
        & \quad - i \tau_1^\beta \tau_2^\gamma
        \left( D_\mu \phi_D {\partial^2 V_{\beta\gamma} \over
        \partial \phi_A \partial \phi_D}
        + D_\mu \psi_\delta
        {\partial^2 V_{\beta\gamma} \over
        \partial \phi_A \partial \psi_\delta} \right), \cr
[ \iso( c_1, \tau_1 ), \iso(c_2, \tau_2 ) ] \xi_\mu{}^\alpha
        & = \iso(c_3, \tau_3) \xi_\mu{}^\alpha
        + i c_1^B c_2^C \left( D_\mu \phi_D
        {\partial^2 W_{BC} \over \partial \psi_\alpha \partial \phi_D}
        + D_\mu \psi_\delta
        {\partial^2 W_{BC} \over \partial \psi_\delta
         \partial \psi_\alpha} \right) \cr
        & \quad - ( c_1^B \tau_2^\gamma - c_2^B \tau_1^\gamma )
        \left( D_\mu \phi_D {\partial^2 U_{B\gamma} \over
        \partial \psi_\alpha \partial \phi_D}
        + D_\mu \psi_\delta
        {\partial^2 U_{B\gamma} \over \partial \psi_\delta
        \partial \psi_\alpha} \right) \cr
        & \quad - \tau_1^\beta \tau_2^\gamma
        \left( D_\mu \phi_D {\partial^2 V_{\beta\gamma} \over
        \partial \psi_\alpha \partial \phi_D}
        + D_\mu \psi_\delta
        {\partial^2 V_{\beta\gamma} \over
        \partial \psi_\delta \partial \psi_\alpha} \right), \cr
  [\iso (c_1, \tau_1), \iso (c_2, \tau_2)]\phi _A
& = \iso (c_3, \tau_3) \phi _A, \cr
 [\iso (c_1, \tau_1), \iso (c_2, \tau_2)]\psi _\alpha
& = \iso (c_3, \tau_3) \psi _\alpha, \cr
}\eqn\SUFINGA
$$
where we have defined
$$\eqalign{
D_\mu \phi_A & \equiv \partial_\mu \phi_A
+ W_{AB} h_\mu{}^B + i U_{A\beta} \xi_\mu{}^\beta, \cr
D_\mu \psi_\alpha & \equiv \partial_\mu \psi_\alpha
- U_{B\alpha} h_\mu{}^B - V_{\alpha\beta} \xi_\mu{}^\beta, \cr
}\eqn\DEFCOV
$$
and the composite parameters
$c_3$ and $\tau_3$ become
$$\eqalign{
c_3^D & \equiv c_1^B c_2^C {\partial W_{BC} \over \partial \phi_D }
- i ( c_1^B \tau_2^\gamma - c_2^B \tau_1^\gamma)
{ \partial U_{B\gamma} \over \partial \phi_D }
+ i \tau_1^\beta \tau_2^\gamma {\partial V_{\beta\gamma} \over
\partial \phi_D}, \cr
\tau_3^\delta & \equiv - i c_1^B c_2^C
{\partial W_{BC} \over \partial \psi_\delta }
+ ( c_1^B \tau_2^\gamma - c_2^B \tau_1^\gamma)
{ \partial U_{B\gamma} \over \partial \psi_\delta }
+ \tau_1^\beta \tau_2^\gamma {\partial V_{\beta\gamma} \over
\partial \psi_\delta}. \cr
}\eqn\COMPP
$$

$D_\mu \phi_A $ and $D_\mu \psi_\alpha $ transform as
coadjoint vectors, i.~e.
$$\eqalign{
 \iso(c, \tau) ( D_\mu \phi_A )
& = c^B \biggl[ {\partial W_{BA} \over \partial \phi_C }
D_\mu \phi_C - {\partial W_{BA} \over \partial \psi_\gamma }
D_\mu \psi_\gamma \biggr]
+ i \tau^\beta \biggl[ {\partial U_{A\beta} \over \partial \phi_C }
D_\mu \phi_C + {\partial U_{A\beta} \over \partial \psi_\gamma }
D_\mu \psi_\gamma \biggr], \cr
 \iso(c, \tau) ( D_\mu \psi_\alpha )
& = c^B \biggl[ {\partial U_{B\alpha} \over \partial \phi_C }
D_\mu \phi_C + {\partial U_{B\alpha} \over \partial \psi_\gamma }
D_\mu \psi_\gamma \biggr]
+  \tau^\beta \biggl[ {\partial V_{\beta\alpha} \over \partial \phi_C }
D_\mu \phi_C - {\partial V_{\beta\alpha} \over \partial \psi_\gamma }
D_\mu \psi_\gamma \biggr]. \cr
}\eqn\DFTRA
$$
Thus $D_\mu$ is recognized as a covariant differentiation.

The commutators of two covariant differentiations
provide
$$\eqalign{
& [ D_\mu, D_\nu ] \phi_A
= W_{AB} R_{\mu\nu}^B + i U_{A\beta} R^\beta_{\mu\nu}, \cr
& [ D_\mu, D_\nu ] \psi_\alpha
= - U_{B\alpha} R_{\mu\nu}^B - V_{\beta\alpha} R^\beta_{\mu\nu}, \cr
}\eqn\DEFCUR
$$
where the curvatures $R_{\mu \nu}^A$ and $R_{\mu\nu}^\alpha$
are defined by
$$\eqalign{
& R_{\mu\nu}^A = \partial_\mu h_\nu{}^A
  - \partial_\nu h_\mu{}^A + {\partial W_{BC} \over \partial \phi_A }
  h_\mu{}^B h_\nu{}^C + i {\partial U_{B\gamma} \over \partial \phi_A }
  ( h_\mu{}^B \xi_\nu{}^\gamma - h_\nu{}^B \xi_\mu{}^\gamma )
  + i {\partial V_{\beta\gamma} \over \partial \phi_A} \xi_\mu{}^\beta
  \xi_\nu{}^\gamma, \cr
& R_{\mu\nu}^\alpha =
\partial_\mu \xi_\nu{}^\alpha
  - \partial_\nu \xi_\mu{}^\alpha - i {\partial W_{BC} \over \partial
  \psi_\alpha }
  h_\mu{}^B h_\nu{}^C + {\partial U_{B\gamma} \over \partial \psi_\alpha }
  ( h_\mu{}^B \xi_\nu{}^\gamma - h_\nu{}^B \xi_\mu{}^\gamma )
  + {\partial V_{\beta\gamma} \over \partial \psi_\alpha} \xi_\mu{}^\beta
  \xi_\nu{}^\gamma, \cr
}\eqn\CURVAT
$$
and the covariant differentiations on $D_\mu \phi _A$ and
$D_\mu \psi_\alpha$ have been
determined in such a way that
the resultant expressions $D_\nu (D_\mu \phi _A)$
and $D_\nu (D_\mu \psi_\alpha)$ transform
like coadjoint fields.
The curvatures do not transform homogeneously.

%
\section{An Invariant action}

We do not have a straightforward way of constructing
invariant actions under the gauge transformation \SUAPHITR,
as noticed from the inhomogeneous form of the transforms
of the curvatures.
Hence by similar discussion to chapter 2,
we will instead seek an action which provides
covariant equations of motion under those transformations,
and examine whether it is invariant under them.

The gauge algebra \SUFINGA\ should be closed,
at least, on shell so as to originate from some Lagrangian symmetry.
Hence the choice
$$
  D_\mu \phi_A = 0,
  \quad D_\mu \psi_\alpha = 0,
\eqn\FEOM
$$
seems an immediate guess for appropriate equations of motion,
which are indeed covariant \DFTRA\ under the transformation
\SUAPHITR.
If $W_{AB}$,
$U_{A\beta}$ and $V_{\beta\alpha}$
are nondegenerate, the equations
\FEOM\ imply
$$
 R_{\mu\nu}^A = 0,
 \quad  R^\alpha_{\mu\nu} = 0,
\eqn\EQNCUR
$$
under the formula \DEFCUR.

The Lagrangian is
constructed from our basic fields
which are the vector ones $h_\mu^A$ and $\xi_\mu{}^\alpha$
the scalar ones $\phi_A$ and $\psi_\alpha$.
Moreover since it is natural that the desired Lagrangian
produces \FEOM\ and \EQNCUR\ as the equations of motion,
we are led to the Lagrangian
$$\eqalign{
 {\cal L} & =
- {1 \over 2} \chi^{\mu\nu}
\biggl[ \phi_A R_{\mu\nu}^A +  i \psi_\alpha R_{\mu\nu}^\alpha
+ \biggl( W_{BC} - \phi_A {\partial W_{BC} \over \partial \phi_A }
- \psi_\alpha {\partial W_{BC} \over \partial \psi_\alpha} \biggr)
h_\mu{}^B h_\nu{}^C  \cr
& \quad
+ 2 i \biggl( U_{B\gamma} - \phi_A {\partial U_{B\gamma}
\over \partial \phi_A }
- \psi_\alpha {\partial U_{B\gamma} \over \partial \psi_\alpha} \biggr)
h_\mu{}^B \xi_\nu{}^\gamma
+ i \biggl( V_{\beta\gamma} - \phi_A {\partial V_{\beta\gamma}
\over \partial \phi_A }
- \psi_\alpha {\partial V_{\beta\gamma} \over
\partial \psi_\alpha} \biggr) \xi_\mu{}^\beta \xi_\nu{}^\gamma \biggr], \cr
}\eqn\LAGCUR
$$
which indeed yields the equations of motion \FEOM\
and \EQNCUR, where $\chi ^{\mu \nu } = -\chi ^{\nu \mu }$.

If we require Poincar{\' e} invariance of the Lagrangian,
the antisymmetric tensor $\chi ^{\mu \nu }$ should be an invariant one
$$
  \chi ^{\mu \nu } = \epsilon ^{\mu \nu },
 \eqn\EPSILON
$$
where $\epsilon ^{\mu \nu }$ denotes
the Levi-Civita tensor in two dimensions.
At this point, we are led to considering the two-dimensional situation.

In summary, we have obtained a candidate action
$$\eqalign{
S & = \int d^2 x \, {\cal L}, \cr
{\cal L} & =
- {1 \over 2} \epsilon^{\mu\nu}
\biggl[ \phi_A R_{\mu\nu}^A +  i \psi_\alpha R_{\mu\nu}^\alpha
+ \biggl( W_{BC} - \phi_A {\partial W_{BC} \over \partial \phi_A }
- \psi_\alpha {\partial W_{BC} \over \partial \psi_\alpha} \biggr)
h_\mu{}^B h_\nu{}^C  \cr
& \quad
+ i \biggl( U_{B\gamma} - \phi_A {\partial U_{B\gamma}
\over \partial \phi_A }
- \psi_\alpha {\partial U_{B\gamma} \over \partial \psi_\alpha} \biggr)
( h_\mu{}^B \xi_\nu{}^\gamma - h_\nu{}^B \xi_\mu{}^\gamma ) \cr
& \quad
+ i \biggl( V_{\beta\gamma} - \phi_A {\partial V_{\beta\gamma}
\over \partial \phi_A }
- \psi_\alpha {\partial V_{\beta\gamma} \over
\partial \psi_\alpha} \biggr) \xi_\mu{}^\beta \xi_\nu{}^\gamma \biggr], \cr
& = - {1 \over 2} \epsilon^{\mu\nu}
[ \phi_A ( \partial_\mu h_\nu{}^A - \partial_\nu h_\mu{}^A )
+ W_{BC} h_\mu{}^B h_\nu{}^C ] \cr
& \quad - {1 \over 2} \epsilon^{\mu\nu}
[ i \psi_\alpha ( \partial_\mu \xi_\nu{}^\alpha
- \partial_\nu \xi_\mu{}^\alpha )
+ i U_{B\gamma} ( h_\mu{}^B \xi_\nu{}^\gamma - h_\nu{}^B \xi_\mu{}^\gamma )
+ i V_{\beta\gamma} \xi_\mu{}^\beta \xi_\nu{}^\gamma ], \cr
}\eqn\SULAGL
$$
which comes out to be diffeomorphism invariant.
The relation \SUJACPHI\ enables us to show
$$\eqalign{
\iso(c, 0 ) {\cal L} & = - \partial_\mu \biggl[ \epsilon^{\mu\nu}
c^B \biggl\{ W_{BC} h_\nu{}^C
+ i U_{B\gamma} \xi_\nu{}^\gamma
- \phi_A \left( {\partial W_{BC} \over \partial \phi_A}
h_\nu{}^C + i {\partial U_{B\gamma} \over
\partial \phi_A } \xi_\nu{}^\gamma \right) \cr
& \quad - \psi_\alpha \left( {\partial W_{BC} \over
\partial \psi_\alpha } h_\nu{}^C + i {\partial U_{B\gamma} \over
\partial \psi_\alpha } \xi_\nu{}^\gamma \right) \biggr\} \biggr], \cr
\iso(0, \tau) {\cal L} & = - \partial_\mu \biggl[ i \epsilon^{\mu\nu}
\tau^\beta  \biggl\{ U_{C\beta} h_\nu{}^C
+ V_{\beta\gamma} \xi_\nu{}^\gamma
- \phi_A \left( {\partial U_{C\beta} \over \partial \phi_A}
h_\nu{}^C + {\partial V_{\beta\gamma} \over
\partial \phi_A } \xi_\nu{}^\gamma \right) \cr
& \quad - \psi_\alpha \left( {\partial U_{C\beta} \over
\partial \psi_\alpha } h_\nu{}^C + {\partial V_{\beta\gamma} \over
\partial \psi_\alpha } \xi_\nu{}^\gamma \right) \biggr\} \biggr], \cr
}\eqn\LAGTR
$$
which confirms the invariance of our action \SULAGL\ under
the `nonlinear' gauge transformation \SUAPHITR.
The equations of motion which follow from \SULAGL\ are given
as follows:
$$\eqalign{
& \epsilon^{\mu\nu} D_\nu \phi_A = 0,
\quad \epsilon^{\mu\nu} R_{\mu\nu}^A = 0, \cr
& \epsilon^{\mu\nu} D_\nu \psi_\alpha = 0,
\quad  \epsilon^{\mu\nu} R_{\mu\nu}^\alpha = 0, \cr
} \eqn\EQMO
$$
which are indeed covariant
due to the transformation law \DFTRA, \TRACUR.

%
\section{Generic Form of `Dilaton' Supergravity}

As a typical example of nonlinear supergauge theory defined in
previous section,
we consider generic form of
two-dimensional $N=1$ dilaton supergravity.
$N=1$ dilaton supergravity has been defined in Ref.~[\Noj].

The superfield formalism are very useful when
we construct a supersymmetric field theory\rlap.\refmark{\Fay}
Also in this paper, we construct $N=1$ dilaton
supergravity by that method.
Thus we introduce superspace $(x^\mu, \theta^\alpha)$,
where $\theta$ is Grassmann number and
$\mu$ and $\alpha$ run over $0$ and $1$.

In order to construct $N = 1$ supersymmetric
extension of two-dimensional dilaton gravity,
we introduce a scalar superfield $\Phi$
and a superzweibein $E_\mu{}^a$ as
$$\eqalign{
& \Phi \equiv \varphi + i \bt \chi + { i \over 2} \bt \theta F, \cr
& E_\mu{}^a \equiv e_\mu{}^a + i \bt \gamma^a \zeta_\mu
        + {i \over 4} \bt \theta e_\mu{}^a A, \cr
}$$
where $\theta$ and ${\bar \theta}$ are real Grassmann numbers
and $\chi$ and $\zeta_\mu$ are fermions.
Since $\theta$ and ${\bar \theta}$ are real,
$\theta$ term is not needed.
Generic form of dilaton supergravity considered in this paper
is the following action: \refmark{\How} \refmark{\Noj}
$$
S_G = 2 i \int d^2 x \, d^2 \theta \, E
	[ \Phi S - {\calf} ( \Phi) ],
\eqn\SUACTION
$$
where
${\calf}( \Phi )$ is a smooth function of $\Phi$ and
$$\eqalign{
& E = e \left[ 1 + {i \over 2} \bt \gamma^\mu \zeta_\mu +
	\bt \theta \left( { i \over 4} A + { 1 \over 8} e^{\mu\nu}
	\bar\zeta_\mu \gamma_5 \zeta_\nu \right) \right], \cr
& S = A + i \bt \Sigma + { i \over 2} \bt \theta G, \cr
& \Sigma = - 2 e^{\mu\nu} \gamma_5 D_\mu \zeta_\nu
        - {1 \over 2} \gamma^\mu \zeta_\mu A, \cr
& G = - R - { i \over 2} {\bar\zeta}_\mu \gamma^\mu \Sigma
	+ { i \over 4} e^{\mu\nu} {\bar\zeta}_\mu \gamma_5
	\zeta_\nu A - {1 \over 2} A^2, \cr
  %
& E_\mu{}^a = e_\mu{}^a + i \bar\theta \gamma^a \zeta_\mu
	+ { i \over 4} \bt \theta e_\mu{}^a A, \cr
& E_\mu{}^\alpha = { 1 \over 2} \zeta_\mu{}^\alpha
	- {1 \over 2}( \bt \gamma_5)^\alpha \omega_\mu
	-{1 \over 4} ( \bt \gamma_\mu )^\alpha A
	- \bt \theta \left( {3i \over 16} \zeta_\mu{}^\alpha A
	+ {i \over 4} ( \gamma_\mu \Sigma )^\alpha \right), \cr
& E_\beta{}^a = i ( \bt \gamma^a )_\beta, \cr
& E_\beta{}^\alpha = \delta_\beta{}^\alpha \left( 1 - {i \over 8}
	\bt \theta A \right). \cr
}\eqn\DEFTWO
$$
Here we denote
$$\eqalign{
& e^{\mu\nu} = e^{-1} \epsilon^{\mu\nu}, \cr
& R = 2 e^{\mu\nu} \partial_\mu \omega_\nu, \cr
& D_\mu \zeta_\nu = \partial_\mu \zeta_\nu + {1 \over 2}
	\omega_\mu \gamma_5 \zeta_\nu, \cr
& \omega_\mu = e_\mu{}^a e^{\nu\rho} \partial_\nu e_{\rho a}
	- { i \over 2} \bar\zeta_\mu \gamma_5 \gamma^\nu \zeta_\nu. \cr
}\eqn\DEFTHR
$$
and $\epsilon^{\mu\nu}$ is the Levi-Civita
antisymmetric tensor,
$a, b$ are tangent space indices,
$\mu, \nu$ are spacetime indices and
$\alpha, \beta$ are spinor indices.
Moreover
we set $\gamma_5 = \gamma^0 \gamma^1$.

The locally well-defined field redefinition
enables us to have \SUACTION\ from the action in Ref.~[\Noj].
The action \SUACTION\ is invariant under the general coordinate,
local-Lorentz and the following local supersymmetry transformation:
$$\eqalign{
& \susy \varphi = i {\bar\epsilon} \chi, \cr
& \susy \chi = [ \gamma^\mu ( \partial_\mu \varphi
	- {i \over 2} {\bar\zeta}_\mu \chi ) - F ] \epsilon, \cr
& \susy F = i {\bar\epsilon} \gamma^\mu \biggl[
	- ( \partial_\mu + { 1 \over 2}
	\omega_\mu \gamma_5 ) \chi + {1 \over 2} \gamma^\nu
	( \partial_\nu \varphi - {i \over 2} {\bar\zeta}_\nu \chi ) \zeta_\mu
	-{ 1 \over 2} F \zeta_\mu \biggr], \cr
& \susy e_\mu{}^a = i {\bar\epsilon} \gamma^a \zeta_\mu, \cr
& \susy \zeta_\mu = 2 ( \partial_\mu + { 1 \over 2} \omega_\mu \gamma_5)
\epsilon + {1 \over 2} \gamma_\mu A \epsilon, \cr
& \susy A = i {\bar\epsilon} \Sigma, \cr
}\eqn\SUSYTRA
$$
where $\epsilon$ is a gauge parameter.

In the following section, we investigate
a gauge-theoretical origin of the `dilaton' superfield $\Phi$
in the action \SUACTION.

%
\section{Nonlinear Super-Poincar{\' e} Algebra}
%


In order to construct the action \SUACTION\ as the nonlinear supergauge
theory, we expand each field in terms of $\theta$ and ${\bar \theta}$,
and carry out the Grassmann integration of
 the action \SUACTION.
Then \SUACTION\ becomes as follows:
$$\eqalign{
S_G & = \int d^2 x \, \biggl[ -{1 \over 2} \epsilon^{\mu\nu}
\biggl\{ \varphi(\partial_\mu \omega_\nu - \partial_\nu \omega_\mu )
- i {\bar\chi} \gamma_5 (\partial_\mu \zeta_\nu
- \partial_\nu \zeta_\mu )
- i \omega_\mu ( {\bar\chi} \zeta_\nu )
+ {1 \over 2} \epsilon_{ab} e_\mu{}^a e_\nu{}^b F A
\biggr\} \cr
& \quad + {1 \over 2} e \{ i {\bar\chi} \chi \wo ( \varphi)
- F \wt (\varphi) \}
+ {i \over 4} \epsilon^{\mu\nu} \epsilon_{ab} e_\mu{}^a
( {\bar\zeta}_\nu \gamma^b \chi) \, \cu (\varphi)
- {1 \over 4} \bigl( e A - {i \over 2} \epsilon^{\mu\nu}
{\bar\zeta}_\mu \gamma_5 \zeta_\nu \bigr) \cv(\varphi) \biggr], \cr
}\eqn\SUGEXP
$$
where we have written ${\calf}(\Phi)$
as
$$
{\calf} (\Phi) = \cv (\varphi) + i \bt \chi \cdot \cu (\varphi)
        + {i \over 2} \bt \theta [ -i {\bar\chi} \chi \cdot
	\wo (\varphi)
        + F \wt (\varphi) ],
$$
$\cv$, $\cu$ and $\wo$ being smooth functions of $\varphi$.

Next we regard the zweibein $e_\mu{}^a$ and
the spinconnection $\omega_\mu$ as the independent fields
as the bosonic case\rlap, \refmark{\Dil}
by which the following term is added to
the action:
$$\eqalign{
S_A & = -{1 \over 2} \epsilon^{\mu\nu} \int d^2 x
         \phi_a T_{\mu\nu}^a, \cr
T_{\mu\nu}{}^a & = \partial_\mu e_\nu{}^a
+ \omega_\mu \epsilon^{ab} e_{\nu b}
+ {i \over 4} \epsilon^a{}_b ( {\bar\zeta}_\mu
\gamma_5 \gamma^b \zeta_\nu ) - ( \mu \leftrightarrow \nu). \cr
}$$
That is, if the auxiliary field $\phi_a $
is integrated out, we go back to the action \SUGEXP\
together with the last equation of
\DEFTHR.

Moreover since $A$ and $F$ are auxiliary fields, they are
eliminated from the action \SUGEXP\ by using the equations of motion
of $A$ and $F$:
$$\eqalign{
& A = \wt (\varphi), \cr
& F = {1 \over 2} \cv (\varphi). \cr
}\eqn\EQNAF
$$

The final expression for the action is as follows:
$$\eqalign{
S_P & = \int d^2 x\, {\cal L}_P \cr
{\cal L}_P & = - {1 \over 2} \epsilon^{\mu\nu}
\biggl[ \varphi ( \partial_\mu \omega_\nu - \partial_\nu \omega_\mu)
- i {\bar\chi} \gamma_5
(\partial_\mu \zeta_\nu - \partial_\nu \zeta_\mu)
+ \phi_a ( \partial_\mu e_\nu{}^a - \partial_\nu e_\mu{}^a )  \cr
& \quad - \left\{ {1 \over 4}  \wt (\varphi) \cv (\varphi)
- i {\bar\chi} \chi \wo (\varphi) \right\}
\epsilon_{ab} e_\mu{}^a e_\nu{}^b
+ 2 \omega_\mu ( \phi_a \epsilon^{ab} ) e_\nu{}_b
\cr
& \quad - i {\bar\chi} \omega_\mu \zeta_\nu
+ {i \over 2} {\bar\chi} \gamma^a \epsilon_{ab} \cu
e_\mu{}^b \zeta_\nu \cr
& \quad - {i \over 2}  {\bar\zeta}_\mu \left\{ {1 \over 2}
\gamma_5 \cv (\varphi)
- \phi_a \epsilon^a{}_b \gamma_5 \gamma^b \right\} \zeta_\nu \biggr]. \cr
}\eqn\SUTOPOL
$$
This action \SUTOPOL\ is classically equivalent to \SUACTION.
Comparing  \SUTOPOL\ with \SULAGL, we find the base nonlinear superalgebra
has the following commutation relation:
$$
\eqalign{
& [ J, P_a ] = \epsilon_{ab} \eta^{bc} P_c, \cr
& [ P_a, P_b ] = - {1 \over 4} \epsilon_{ab} \wt (J) \cv (J)
        - {i \over 2} \epsilon_{ab} (Q \gamma^0 Q) \wo (J), \cr
& [J, Q_\alpha ] = {1 \over 2} ( Q \gamma_5 )_\alpha, \cr
& [ P_a, Q_\beta ] = { 1 \over 4} ( Q \gamma^0
\gamma^b \gamma^1 \epsilon_{ba} )_\alpha \cu (J), \cr
& \{ Q_\alpha, Q_\beta \} =
	{i \over 2} (\gamma^a \gamma^0 )_{\alpha\beta} P_a
        + {i \over 4} ( \gamma^1 )_{\alpha\beta} \cv (J). \cr
} \eqn\SUMODISO
$$
and the other commutation relations vanish.
Here we set
$\{ T_A \} = \{ P_a, J \}$, that is, $T_0 = P_0$, $T_1 = P_1$,
and $T_2 = J$, and
we also set the vector fields $h_\mu^A = (e_\mu{}^a, \omega_\mu)$,
$\xi_\mu{}^\alpha = ( \gamma^1 \zeta_\mu{} )^\alpha $ and
the scalar field $\phi_A = (\phi_a, \varphi)$,
$\psi_\alpha = {\chi^{\dag}}_\alpha$;
$\eta^{cd}$ is the two-dimensional Minkowski metric.

Note that the choice $\wo (J) = \cv(J) = \cu(J) = 0$
corresponds to the original super-Poincar{\' e} algebra.

Then the structure functions defined in \NLLALG\
for the above algebra are given by
$$
\eqalign{
& W_{2a} = - W_{a2} = \epsilon_a{}^b \phi_b,
\qquad W_{22} = 0, \cr
& W_{ab} = - {1 \over 4} \epsilon_{ab} \wt (\varphi) \cv (\varphi)
        + {i \over 2} \epsilon_{ab} {\bar\chi} \chi \wo (\varphi), \cr
& U_{2\beta} = - { 1 \over 2} {\bar\chi}_\beta, \cr
& U_{a\beta} = {1 \over 4} ( {\bar\chi} \gamma^b
\gamma^1 )_\beta \epsilon_{ba} \cu (\varphi), \cr
& V_{\alpha\beta} = - {1 \over 4} (\gamma^0 \gamma_5 )_{\alpha\beta}
        \cv (\varphi) - {1 \over 2} \phi_a
        ( \gamma^a \gamma^0 )_{\alpha\beta}. \cr
} \eqn\HATENA
$$

The gauge transformation law \SUAPHITR\
now reads
$$
 \eqalign{
  \iso \omega_\mu \, &= \partial_\mu t
                    + \epsilon_{bc} c^b e_\mu{}^c
               {\partial \over \partial \varphi} \biggl[ {1 \over 4}
	\wt (\varphi)\cv (\varphi) - {i \over 2} {\bar\chi} \chi
	\wo (\varphi) \biggr] \cr
	& \quad + {i \over 4} \epsilon_{bc} c^b
	( {\bar\chi} \gamma^c \zeta_\mu )
	{\partial \cu \over \partial \varphi }
	+ {i \over 4}  ( {\bar\chi} \gamma^b \tau) \epsilon_{bc} e_\mu{}^c
	 {\partial \cu \over \partial \varphi }
	+ {i \over 4} ( {\bar\tau} \gamma_5 \zeta_\mu )
	{ \partial \cv \over \partial \varphi}, \cr
  \iso e_\mu{}^a &= - t \, \epsilon^{ab} e_{\mu b}
                 + \partial_\mu c^a + \omega_\mu \epsilon^{ab} c_b
		- {i \over 2} \epsilon^{ab} ( {\bar\tau} \gamma_5
		\gamma_b \zeta_\mu ), \cr
  \iso \zeta_\mu{}^\alpha &= - {1 \over 2} t ( \gamma_5 \zeta_\mu)^\alpha
	+ \epsilon_{bc} c^b e_\mu{}^c (\gamma_5 \chi)^\alpha \wo(\varphi)
	- {1 \over 4} \epsilon_{bc} c^b ( \gamma_5 \gamma^c
	\zeta_\mu)^\alpha \cu(\varphi) \cr
	& \quad + \partial_\mu \tau^\alpha
	- {1 \over 4} ( \gamma_5 \gamma^b \tau )^\alpha
	\epsilon_{bc} e_\mu{}^c \cu(\varphi), \cr
  \iso \varphi \,\, &= \epsilon^{ab} c_a \phi_b
	+ {i \over 2} {\bar\chi} \tau, \cr
  \iso \phi_a &= - t \epsilon_{ab} \phi^b
                + \epsilon_{ab} c^b \biggl[{1 \over 4} \wt(\varphi)
		\cv (\varphi)
		- {i \over 2} {\bar\chi} \chi \wo(\varphi) \biggr], \cr
  \iso \chi^\alpha &= - {1 \over 2} t (\gamma_5 \chi)^\alpha
	+ {1 \over 4} (\gamma_5 \gamma^a \chi)^\alpha \epsilon_{ab} c^b
	\cu(\varphi)
	- {1 \over 4} \cv (\varphi) \tau^\alpha
	+ {1 \over 2} \phi^a \epsilon_{ab}(\gamma^b \tau)^\alpha, \cr
}
 \eqn\SUMISO
$$
where we have put $c^A = (c^a, t)$.

The equations of motion which follow from \SUTOPOL\ are given by
$$
 \eqalignno{
  &\partial_\mu \varphi + \phi_a \epsilon^{ab} e_{\mu b}
   - {i \over 2} ({\bar\chi} \zeta_\mu) = 0, \cr
  &\partial_\mu \phi_a + \omega_\mu \epsilon_{ab} \phi^b
   + \epsilon_{ab} e_\mu{}^b \biggl( - {1 \over 4} \wt (\varphi) \cv (\varphi)
+ {i \over 2}  {\bar\chi} \chi \wo (\varphi) \biggr)
- {i \over 4} \epsilon_{ab} ( {\bar\chi} \gamma^b \zeta_\nu ) \cu = 0, \cr
  & \partial_\mu {\bar\chi} - {1 \over 2} {\bar\chi} \gamma_5 \omega_\mu
+ {1 \over 4} {\bar\chi} \gamma_5 \gamma^a \epsilon_{ab} e_\mu{}^b \cu
- {1 \over 2} {\bar\zeta}_\mu ( {1 \over 2} \cv -  \gamma^a \epsilon_{ab}
\phi^b ) = 0,\cr
&{1 \over 2} \epsilon^{\mu \nu} \biggl[
\partial_\mu \omega_\nu - \partial_\nu \omega_\mu
+ \epsilon_{bc} e_\mu{}^b e_\nu{}^c {\partial \over \partial \varphi}
\biggl\{ { 1 \over 4} (\wt \cv) + {i \over 2} {\bar\chi} \chi \wo \biggr\}
- {i \over 2} \epsilon_{bc} e_\mu{}^b ( {\bar\chi} \gamma^c \zeta_\nu )
{\partial \cu \over \partial \varphi}
- { i \over 4} {\partial \cv \over \partial \varphi} ( {\bar\zeta}_\mu \gamma_5
\zeta_\nu ) \biggr]
= 0, \cr
& {1 \over 2} \epsilon^{\mu \nu} \biggl[
\partial_\mu \zeta_\nu - \partial_\nu \zeta_\mu
- \epsilon_{bc} e_\mu{}^b e_\nu{}^c \wo \gamma_5 \chi
+ \omega_\mu \gamma_5 \zeta_\nu
+ {1 \over 2} \epsilon_{bc} e_\mu{}^b \cu
 \gamma_5 \gamma^c \zeta_\nu \biggr] = 0, \cr
&    {1 \over 2} \epsilon^{\mu \nu} T_{\mu \nu}{}^a = 0.
& \eqname\EQOM \cr
 }
$$
They of course correspond to \EQMO.

Note that the gravitation theory \SUGEXP\ is obtained
if one adds \EQNAF\ to the Lagrangian \SUTOPOL\ and
integrates out the fields $\phi_a$ and $\omega_\mu$
in the Lagrangian \SUTOPOL\ under the condition $e \neq 0$.
The reformulation \SUTOPOL\ reveals that the theory \SUACTION\ possesses
hidden gauge symmetry \SUMISO\ of the Yang-Mills type.
This gauge symmetry includes diffeomorphism, as is the case
for the Yang-Mills-like formulation
of generic form of dilaton gravity\rlap.
\refmark{\Dil}

%
\chapter{Conclusion}

We have made a systematic
construction of the nonlinear gauge theory \LAGL\
with the gauge transformation \APHITR\ based on
a general nonlinear Lie algebra.
A new category of the gauge theory has been constructed
as an extension
of the gauge theory based on the usual Lie algebra.
We have carried out the BRS quantization of nonlinear gauge theory.
Since the gauge is open,
we have had to modify usual quantization method.
The BRS charge has been explicitly constructed without
anomaly in the cylindrical
spacetime $S^1\!\times{\mib R}^1$ in the temporal gauge.

The algebraic structure of the above nonlinear gauge theory
seems to be clear by construction.
However the field $\Phi_A$ is not a representation of nonlinear Lie algebra;
it is different from the coadjoint field of usual Lie algebra.
The relation of $\Phi_A$ with the algebra
deserves further investigation.

The clarification of its geometric structure
is desirable in view of the rich structure
present in the usual nonabelian gauge theory.
However, the gauge algebra \FINGA\ is open,
which implies non-existence of Lie-group-like object
corresponding to the nonlinear Lie algebra.
That might cause difficulty in the geometrical investigation
of the nonlinear gauge theory.

In the chapters 3 and 4,
we have considered typical examples which have nonlinear
gauge symmetry defined in previous chapter.
We have considered Lorentz-covariant quadratic extension \MODISO\
of the Poincar{\' e} algebra in two dimensions in chapter 3.
The nonlinear gauge theory based on it has turned out to be
the Yang-Mills-like formulation of $R^2$ gravity with
dynamical torsion.
Namely, the ``deformation'' we observed in Ref.[\Ike]
has been shown to be Lorentz-covariant quadratically nonlinear extension
described above.


We have also considered general Lorentz-covariant nonlinear extension \MODISO\
of the
\hfil\break
Poincar{\' e} algebra in two dimensions in chapter 4.
The nonlinear gauge theory based on its algebra has turned out to be
the generic form of `dilaton' gravity,
which clarifies a gauge-theoretical origin of the non-geometric
scalar field in two-dimensional gravitation theory.
We note that this theory is a generalization
of Yang-Mills-like formulation of
the Jackiw-Teitelboim's model\refmark{\Fuk}
and the dilaton gravity\rlap,\refmark{\Ver}
because the latter corresponds to a particular choice
${\cal W}(\varphi) = {\rm constant}$.

%
%

In chapter 5, we have made a
construction of the nonlinear supergauge theory \SULAGL\
with the gauge transformation \SUAPHITR\ based on
a general nonlinear Lie superalgebra.


We have also found that
the generic form of `dilaton' supergravity
turns out to be the nonlinear supergauge theory based on the
Lorentz-covariant nonlinear extension \SUMODISO\
of the super-Poincar{\'e} algebra in two dimensions.
The reformulation \SUTOPOL\ reveals that the theory \SUACTION\ possesses
hidden gauge symmetry \SUMISO\ of the Yang-Mills type.
We also can interpret two formulations
of dilaton supergravity
that the super structure of the base space
in dilaton supergravity is transmuted into
the super structure of target space
in a nonlinear super-Poincar\'e algebra.
This theory is a generalization
of Yang-Mills-like formulation of the dilaton supergravity\rlap.
\refmark{\Riv}


We expect that construction of the nonlinear gauge theory
and its application to two-dimensional gravity
in this paper has offered
a new method to analyze quantum field theory.
The analysis of this theory at the quantum level
is still insufficient.
It might be interesting to compare
the theory \SUACTION\ of the Utiyama type
and the one \SUTOPOL\ of the Yang-Mills type in the aspects of their
quantization.

%
\acknowledge
The author would like to thank N.~Nakanishi
for valuable comments and careful reading of the manuscript.
He thanks K.~-I.~Izawa and for valuable discussion as his collaborator.
He also thanks I.~Ojima and M.~Abe for valuable
comments and discussions.
%


\endpage

\refout

\bye